\input{style/arxiv-ba.cfg}
\documentclass[ba,linksfromyear,preprint]{imsart}
\makeatletter
   \@ifpackageloaded{natbib}{}{\usepackage{natbib}}
\makeatother
\usepackage{bm}
\usepackage{multirow}

\pubyear{2015}
\volume{10}
\issue{2}
\firstpage{379}
\lastpage{410}
\doi{10.1214/14-BA919}

\begin{document}

\begin{frontmatter}
\title{Predictions Based on the Clustering of Heterogeneous Functions
via Shape and Subject-Specific Covariates}
\runtitle{Prediction via Curve Clustering}

\begin{aug}
\author[a]{\fnms{Garritt L.} \snm{Page}\ead[label=e1]{page@mat.puc.cl}}
\and
\author[b]{\fnms{Fernando A.} \snm{Quintana}\ead[label=e2]{quintana@mat.puc.cl}}

\runauthor{G. L. Page and F. A. Quintana}

\address[a]{Departmento de Estad\'istica, Pontificia Universidad Cat\'olica
de Chile, Santiago, Chile,\\\printead{e1}}
\address[b]{Departmento de Estad\'istica, Pontificia Universidad Cat\'olica
de Chile, Santiago, Chile,\\\printead{e2}}

\end{aug}

%
\begin{abstract}
We consider a study of players employed by teams who are members of the
National Basketball Association where units of observation are
functional curves that are realizations of production measurements
taken through the course of one's career. The observed functional
output displays large amounts of between player heterogeneity in the
sense that some individuals produce curves that are fairly smooth while
others are (much) more erratic. We argue that this variability in curve
shape is a feature that can be exploited to guide decision making,
learn about processes under study and improve prediction. In this paper
we develop a methodology that takes advantage of this feature when
clustering functional curves. Individual curves are flexibly modeled
using Bayesian penalized B-splines while a hierarchical structure
allows the clustering to be guided by the smoothness of individual
curves. In a sense, the hierarchical structure balances the desire to
fit individual curves well while still producing meaningful clusters
that are used to guide prediction. We seamlessly incorporate available
covariate information to guide the clustering of curves
non-parametrically through the use of a product partition model prior
for a random partition of individuals. Clustering based on curve
smoothness and subject-specific covariate information is particularly
important in carrying out the two types of predictions that are of
interest, those that complete a partially observed curve from an active
player, and those that predict the entire career curve for a player yet
to play in the National Basketball Association.
\end{abstract}

%


\end{frontmatter}


\section{Introduction}

In multi-subject studies where observations are considered to be
functional realizations, it is common to observe large amounts of
between-subject heterogeneity in the sense that some subjects produce
curves that are quite smooth while others produce curves that are
(much) more erratic. Although not often explicitly considered when
modeling these types of data, the variability in curve shape can be an
important feature that may help distinguish individuals and lead to
better understanding of processes under study and/or better
predictions. Often in these studies two types of predictions are
desired: those that predict the functional output across the entire
time domain for a hypothetical new subject and those that complete a
partially observed functional response for subjects currently
participating in the study. Explicitly considering curve shape in
modeling, in addition to considering relevant covariates, should
improve both types of predictions. This is particularly true of the
application motivating the present study. Decision makers of teams that
belong to the National Basketball Association (NBA) are very interested
in being able to group and predict future performance of basketball
players that are employed or could possibly be employed by NBA teams.

%
\begin{figure}[htbp]
\includegraphics{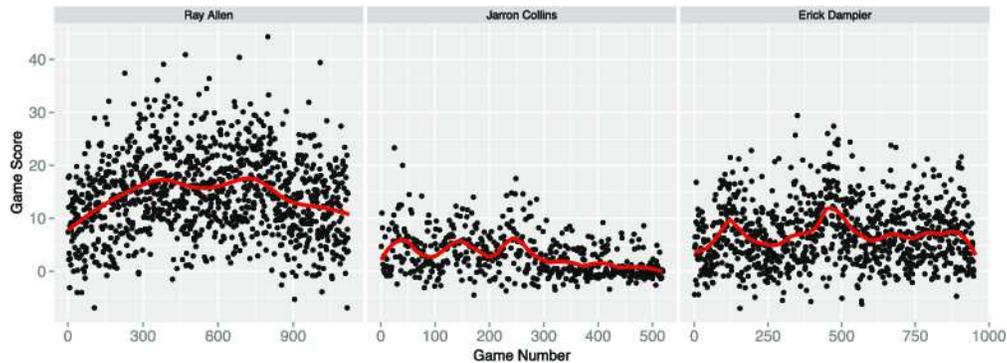}
\caption{Career game-by-game Game Score results for three NBA players.
A loess smoother with span equal to 0.3 is also provided.}
\label{rawscatterplot}
\end{figure}

The NBA is a North American professional mens basketball league that
arguably employs the worlds most gifted basketball players. The
league's popularity has steadily increased and as a result player
salaries have exploded. Personnel decisions in the NBA (as in most
professional sports leagues) are high-risk transactions. In the face of
massive amounts uncertainty teams offer players guaranteed multi-year
multi-million dollar contracts and as a result mistakes in player
acquisition are extremely expensive. Making things even more
treacherous are the abstruse rules governing player transactions found
in the collective bargaining agreement (CBA). Among other things, the
CBA regulates the amount of resources dedicated to player acquisition.
Teams that misallocate player salary resources by over paying severely
hinder a team's future flexibility and negatively impact a team's
future competitiveness and profitability for years. Because of this,
added value might be assigned to players who perform consistently
compared to those that are more up and down.

Figure \ref{rawscatterplot} displays scatter-plots and loess curves
(with a span of 0.3) of game-by-game ``production'' for three NBA
basketball players. Game-by-game ``production'' in Figure \ref{rawscatterplot}
is measured using the so called Game Score statistic
(\citealt{hollinger2002pro}). More details are provided in Section \ref{examples}
and the Appendix, but for now it suffices to know that
higher values correspond with better performance and more production.
Even though there is a large amount of game-to-game and
player-to-player variability in Game Score it is still evident that
production consistency between the three players varies. Erick Dampier
appears to have two spikes of improved production, Jarron Collin's
production oscillates during the first part of his career while Ray
Allen's production is fairly smooth as it gradually increases and
decreases with slight dip in the middle. Therefore curve shape should
contain information that is valuable in distinguishing between
different types of players and being able to assess their future value.
Two types of predictions are used to assess future value. The first
considers players who are currently members of the NBA and that will
continue to participate in future games. Ray Allen in Figure \ref
{rawscatterplot} is an example of such a player. Although he has
already played in more than 1000 games, he continues to play and
predicting the performance for the remainder of his career is of
considerable interest. This type of prediction will be referred to as
``active player prediction''. The second type of prediction considers
basketball players who have yet to play in an NBA game but who have a
skill set that will attract interest from NBA teams. For these players,
predicting the entire career production curve is of interest. This type
of prediction will be referred to as ``career prediction''.

It has become fairly common to consider the longitudinal curves of the
type just described as discretized realizations of functional data.
There is now a large literature dedicated to functional data analysis
(FDA) techniques. A few popular methods that are actively being
researched are functional principal components (\citealt[chap.
6]{FunctionalDataAnalysisBook}, \citealt
{MultilevelFunctionalPrincipalComponentAnalysis}), Gaussian process
regression methods (\citealt{rasmussen}, \citealt
{HierModelsAssesVarAmongFunctionsBehesta}, and \citealt
{LocallyNPBRegViaGP}) and multi-level functional basis expansion
(\citealt{WaveletBasedFunctionalMixedModel}, \citealt
{BayesFreeKnotCurveFit}, \citealt{Biller:2000}, \citealt
{BayesianLatentFactorRegressionForFunctionalAndLongitudinalData},
\citealt{AdaptiveRegressionSplines}). When considering multiple-subject
studies the methods just described tend to separate individuals
according to trend levels only, while ignoring the shape of the
longitudinal projections. Though the idea of explicitly using shape or
smoothness of curves to improve prediction is intuitively appealing
there is surprisingly very little in the statistical literature
dedicated to it. The one article that we are aware of is \cite
{StochasticVolatilityRegressionForFunctionalDataDynamics} whose focus
is on estimating rate functions through a complicated system of
differential equations and using covariates to explain variability in
trajectories via stochastic volatility models. They applied their
method to a longitudinal multi-subject blood pressure study for
pregnant women and noted that blood pressure trajectories for normal
women were more smooth relative to women with preeclampsia. We however,
take an entirely different approach. Instead of dealing with a
complicated system of differential equations we incorporate curve shape
in modeling through a penalty parameter analogous to that found in
penalized splines.

Our model involves an implied distribution on partitions of players.
The allocation variables are treated as parameters and thus our
approach may be seen as an extension of latent class analysis (LCA)
(\citealt{LCAbook}) which classifies individual player curves into $K$
pre-specificed clusters (see \citealt{LCA:2010} and \citealt
{LCA:2005}). Unlike LCA, our methodology does not require a fixed
pre-specified number of clusters as this is inferred from the
corresponding posterior distribution on partitions. We briefly note
that there does exist a small literature dedicated to estimating
certain aspects of functional output such as dynamics (the speed of
price increases and the rate at which this speed changes) that depend
on covariate information (see \citealt{EbayFunctions} and \citealt
{RateFunctions} and references therein). But these are not relevant to
the current setting as they fail to deal with multiple-subject studies
nor do they use curve shape in prediction and inference.

In sports, \cite{reese:1999} model career trajectories (or aging
curves) non parametrically in order to make historical comparisons of
player's abilities in baseball, hockey and golf. \cite{StreakyPGAPlay}
consider career paths of golf players to determine combinations of luck
and skill required to win a golf tournament. Neither of these works
were interested in grouping players to carry out career and active
player predictions.

As noted, our principal goal is making active player and career
predictions. If predictions are computed using methods that treat
individual players independently, then both types of predictions would
be extremely poor as they would not be data driven. One way of
improving predictions is by borrowing strength (or sharing information)
among players whose career production curves might be deemed similar. A
straightforward way of borrowing strength is by introducing player
clusters. However, if all individuals of the same cluster are
restricted to have the same curve, then some individuals will
invariably be poorly fit (too much borrowing of strength).
Alternatively, if curves of all individuals of the same cluster are
completely unrestricted, then clustering players would provide no
predictive information (too little borrowing of strength). The
methodology proposed in this article is able to balance very well the
desire to produce good fitting individual curves while still producing
clusters that allow enough borrowing of strength among similar players
to guide prediction. This is carried out by employing a hierarchical
model where subject-specific functions are modeled flexibly through a
linear combination of basis functions whose coefficients are drawn from
cluster-specific process level distributions. Doing this produces
flexible subject-specific curves while still being able to produce
reasonably accurate predictions by pooling together players with
similar features/performances.

In our model, having covariate dependent clusters is crucial to
carrying out career prediction as these are produced using the
predictive distribution available from the covariate dependent
clustering mechanism. Also, shape dependent clusters are useful to
carrying out active player predictions as incomplete active player
curves are filled in using curves of retired players that have similar
career trajectories. There has been work regarding completing curves
(\citealt{PredictingFunctionsCallCenter} work out Best Linear Unbiased
Predictors (BLUPS) for past and future curve segments) and local
borrowing of strength to fit global curves (\citealt
{PetroneGuindaniGelfand} employ functional Dirichlet processes to group
Gaussian process realizations), but the approaches developed and
purposes are very much different from the present study.

The remainder of the article is organized as follows. Section 2
describes the data collected and employed in the analysis. Section 3
provides details regarding the development of the methodology
highlighting model components associated with cluster-specific curve
smoothness and active player prediction. Section 4 provides details
regarding computation of posterior and predictive distributions. In
Section 5 we provide details of a small simulation study. Results from
the analysis of the NBA application are provided in Section 6. Finally,
we provide some concluding remarks in Section 7.

\section{Description of Data} \label{examples}
We collected common game-by-game (including playoffs) modern basketball
metrics for each player drafted into the NBA during the years 1992-2004
(Shaquille O'Neal to Dwight Howard) that participated in at least one
game up through the 2009/2010 season. This resulted in 576 players with
number of games played ranging from 2 to 1383 games. A few of the
players appeared in very few games and are not representative of a
typical NBA player. Because of this, and to reduce the noise introduced
by the careers of players that contain little information regarding the
processes of interest, we restrict our attention to players with at
least three seasons of experience (the rookie contract length of the
2005 CBA). Also, to retain enough games to get a reasonable sense of a
player's ability we only include players who played at least a half a
season's worth of games (42). Finally, we excluded the following 8
players whose careers were cut short either by career ending injuries
or untimely deaths: Bryant Reeves, Malik Sealy, Eddy Curry, Jason
Collier, Eddie Griffin, Yao Ming, T. J. Ford, and Gilbert Arenas. This
resulted in 408 players with number of games played through the
2009/2010 season ranging from 45 to 1383. Of the 408 players, 263 are
classified as ``retired'' as they did not play beyond the 2009/2010 season.

Measuring game-by-game production is not straightforward as there are
numerous, difficult to measure factors that influence player
performance. Because of this no gold standard basketball production
metric exists. That said, one that has become somewhat popular is John
Hollinger's so called Game Score which is a linear combination of
common variables that are recorded for each player through out the
course of a game (e.g., number of baskets made and number of steals
acquired. More details can be found in the Appendix and at \citealt
{hollinger2002pro}). This metric will be used as our response variable
and therefore a representation of a player's game productivity. Though
Game Score has deficiencies (e.g., weighted heavily towards offensive
output and doesn't account for quality of opponent), it provides a
fairly accurate indicator of player production for any given game. The
maximum Game Score collected is 63.5 (corresponding to Kobe Bryant's 81
point game). The minimum Game Score was -9.9 and the average Game Score
among all players is 8.1. An alternative to raw Game Score is a
standardized Game Score where standardization is carried out by
dividing Game Score by the minutes played in each game, thus removing
Games Score's dependence on minutes played. However, players whose
production is not negatively impacted by increased minutes are more
valuable than those who are less efficient with increased game time and
distinguishing between these types of players is desirable. For this
reason we opt to use raw Game Score values.

For an aging (or time) variable there are various units that could be
used. For example, age, number of accumulated minutes played, or simply
the number of games played are all reasonable. Since each of these
measurements are only available to us on a game-by-game basis, the
shape (or smoothness) of the curve remains unchanged regardless of age
unit employed. Thus for sake of expositional clarity we use number of
games played (see Figure \ref{rawscatterplot} as an example).

Through exploratory analysis we identified three covariates that in
addition to being of interest in their own right are informative in
grouping players. These are age during first game played (measured in
years), experience before being drafted into the NBA (High School
basketball only, some Collegiate basketball, or International
basketball), and draft order. Draft order is the order in which players
are selected in their respective drafts. For example, a player with
draft order $1$ implies he was the first player selected in his
respective draft, and draft order $2$ implies he was the second player
selected etc. In Section 3 we describe how draft order is used
explicitly to predict total games played for active players, but as a
covariate used to influence clustering we categorize a player's draft
order as being a top five pick, a first round pick (excluding the first
five) and a second round pick. (Since 1989 the NBA draft has consisted
of two rounds.) Table \ref{playercategorysummary} provides the number
of players in each of the nine categories. Other baseline covariates
were considered such as position, height, and other physiological
characteristics but preliminary research indicated they were not useful
in partitioning players with regards to production.

%
\begin{table}[htdp]
\caption{Total number of players in each of the nine categories.}\label{playercategorysummary}
\vspace*{4pt}
\begin{tabular}{l ccc}
& \multicolumn{3}{c}{Experience} \\ \cline{2-4}
Draft & High School & College & International\\\hline
Top 5 & 7 & 51 & 4 \\
1st Round & 15 & 200 & 25\\
2nd Round & 1 & 98 & 8\\\hline
\end{tabular}
\end{table}
%

\section{Model Description and Development}\label{MODEL}
We first consider the model's clustering mechanism highlighting its
dependence on subject-specific covariates. Secondly, the likelihood
structure incorporating the number of games played and career length
(which are right censored for active players) is detailed. Lastly, we
describe the hierarchical component which balances goodness of
individual fit with ability to produce clusters that are able to guide
prediction.

\subsection{Product Partition Model with Covariates (PPMx)}
Let $i = 1, \ldots, m$ index the $m$ players in the study. Further, let
$\rho= \{S_1, \ldots, S_{k_m}\}$ denote a partitioning (or clustering)
of the $m$ individuals into $k_m$ subsets such that $i \in S_j$ implies
that individual $i$ belongs to cluster $j$. Alternatively, we will
denote cluster membership using $s_1, \ldots, s_m$ where $s_i = j$
implies $i \in S_j$. Let $\bm{x}_i = (x_{i1}, x_{i2}, x_{i3})$ denote
player $i$'s covariate vector with $x_{i1}$ corresponding to age,
$x_{i2}$ experience and $x_{i3}$ draft order. Let $\bm{x}^{\star}_j = \{
\bm{x}_i:i \in S_j\}$ be the partitioned covariate vector. Our approach
is to first directly model $\rho$ with the covariate dependent product
partition model of \cite{PPMxMullerQuintanaRosner} (which will be
referred to as the PPMx model) and then construct a hierarchical model
given the partition (as opposed to introducing latent variables that
indirectly induce a partitioning of individuals). The PPMx prior
incorporates the idea that individuals with similar covariate values
are more likely {\it a priori} to belong to the same cluster relative
to individuals with dissimilar covariate values. Additionally, this
prior is very simple, highly customizable, seamlessly incorporates
different types of covariates (e.g., continuous or categorical), and is
particularly well suited for prediction (something that is of interest
here). An alternative method not considered here can be found in \citet
{BayesianGeneralizedProductPartitionModel}. The PPMx prior consists of
a cohesion function, $c(S_j) \ge0$ for $S_j \subset\{1, \ldots, n\}$,
and a nonnegative similarity function $g(\bm{x}^{\star}_j)$. The former
measures the tightness of how likely elements of $S_j$ are clustered
{\it a priori} and the latter formalizes the similarity of the $x_i$'s
by producing larger values of $g(\bm{x}^{\star}_j)$ for $x_i$'s that
are more similar. The form of the PPMx prior is simply the following
product (for more details see \citealt{PPMxMullerQuintanaRosner})
\begin{align} \label{ppmx}
P(\rho|\bm{x}) \propto\prod_{j=1}^{k_m} c(S_j)g(\bm{x}^{\star}_j).
\end{align}

A simple example of a cohesion function that produces a Dirichlet
Process type partitioning is $c(S_j) = M\times(|S_j| - 1)!$ for some
positive $M$ and $|\cdot|$ denoting cardinality. Regarding possible
similarity functions, \citet{PPMxMullerQuintanaRosner} provide a bit of
exposition for different types of covariates (e.g., continuous,
ordinal, or categorical). Generically speaking they suggest the
following structure
\begin{align}\label{sf}
g(\bm{x}_j^{\star}) = \int\prod_{i \in S_j} q(x_i | \zeta_j) q(\zeta_j)
d\zeta_j.
\end{align}
where $\zeta_j$ is a latent variable and $q(\cdot|\zeta_j)$ and $q(\cdot
)$ are (typically) conjugate probability models. This structure is not
necessarily used for its probabilistic properties (indeed $\bm{x}$ is
not even random), but rather as a means to measure the similarity of
the covariates in cluster $S_j$. In reality any function that produces
larger values as the entries of $\bm{x}_j^{\star}$ become more similar
can be considered as a similarity function. For example $g(\bm{x}^{\star
}_j) = \exp\{ - s^2_j\}$ where $s^2_j$ is the empirical variance of $\bm
{x}^{\star}_j$ is a completely reasonable similarity function for
continuous covariates.

It turns out that the similarity function \eqref{sf} coupled with
cohesion function $c(S_j) = M\times(|S_j| - 1)!$ produces the same
marginal prior distribution on partitions as that induced by using a
Dirichlet process (DP). For more details see \citet{PPMxMullerQuintanaRosner}.

Given $\rho$ we may proceed to specify a hierarchical model that
flexibly models individual curves. Before doing so, we very briefly
introduce a few pieces of notation that will be used. In what follows
cluster-specific and subject-specific parameters will need to be
distinguished. If we let $\bm{\theta}_i$ denote some generic
subject-specific parameter vector, then $\bm{\theta}^*_j$ will be used
to denote a cluster-specific parameter in the sense that $i \in S_j$
implies that $\bm{\theta}_i = \bm{\theta}^*_j$. Alternatively, cluster
labels $(s_1, \ldots, s_m)$ can be used to connect subject and cluster
specific parameters through $\bm{\theta}_i = \bm{\theta}^*_{s_i}$.
Lastly, vectors of subject-specific and cluster-specific parameters are
denoted by $\bm{\theta} = (\bm{\theta}_1, \dots, \bm{\theta}_m)$ and
$\bm{\theta}^* = (\bm{\theta}^*_1, \ldots, \bm{\theta}^*_{k})$.

\subsection{Likelihood}
To distinguish between players that play beyond the 2009/2010 season we
use the following indicator variable\vadjust{\eject}
\begin{align*}
g_i = \left\{
\begin{array}{cl}
0 & \mbox{if player $i$ retired before or at the conclusion of the
2009/2010 season} \\
1 & \mbox{if player $i$ played beyond the 2009/2010 season}.
\end{array}
\right.
\end{align*}
Let $n_i$ denote the total number of games played in player $i$'s
career. If $g_i=0$ then $n_i$ is observed otherwise a lower bound
denoted by $\tilde{n}_i$ is observed such that $n_i \ge\tilde{n}_i$.
Thus, we are dealing with right censored type observations and we will
incorporate ideas developed for modeling them. We denote the response
vector for players whose $g_i = 0$ with $\bm{y}_i = (y_{i1}, \ldots,
y_{in_i})$ otherwise $\bm{y}_i = (y_{i1} \ldots, y_{i\tilde{n}_i})$.
The career production curve for players whose $g_i=1$ needs to be
``completed'' which requires the prediction or imputation of ${n}_i$.

Predicting $n_i$ is not trivial (even given $\tilde{n}_i$) because it
is highly variable and demonstrates a strong association with very few
covariates. One covariate we found that displays a strong association
with $n_i$ is career length (denoted by $L_i$ and measured in years).
Unfortunately, this variable is also right censored and for $g_i=1$ we
only observe $\tilde{L}_i$. However, we consider $L_i$ because it
displayed a stronger association with the uncensored variable draft
order (denoted by $d_i$) than that found between $n_i$ and $d_i$.
Therefore we employ $d_i$ to first impute $L_i$ and then use $L_i$ to
predict $n_i$. Thus, the likelihood for the $i$th player is composed of
the random variables $(\bm{y}_{i}, g_i, n_{i:g_i=0}, \tilde
{n}_{i:g_i=1}, L_{i:g_i=0}, \tilde{L}_{i:g_i=1})$ which we model
jointly by way of
\begin{align*}
p(\bm{y}_{i}, g_i, n_{i:g_i=0}, \tilde{n}_{i:g_i=1}, L_{i:g_i=0}, \tilde
{L}_{i:g_i=1}) & = [p(\bm{y}_{i} | n_i, L_i) p(n_i|L_i)p(L_i)]^{1-g_i}
\\
& \times[p(\bm{y}_{i} | \tilde{n}_i, \tilde{L}_i) p(\tilde{n}_i|\tilde
{L}_i)p(\tilde{L}_i)]^{g_i}.
\end{align*}
We now detail each of the three likelihood components.


When only considering retired players we found that the association
between $d_i$ and $L_i$ was somewhat nonlinear. (This is reasonable
considering that our pool of players consists of only those who played
at least one NBA game thus retaining only the ``good'' 2nd round
picks.) Because of this, we assume $L_i \sim N(\nu_i, \psi^2)$ where
$\nu_i = \gamma_0 + \gamma_1d_i + \gamma_2d_i^2$. However, the
association between $n_i$ and $L_i$ was fairly linear so we assume $n_i
| L_i \sim N(\eta_i, \delta^2)$ where $\eta_i = \alpha_0 + \alpha
_1L_i$. Thus, $(n_{i:g_i=0}, \tilde{n}_{i:g_i=1}, L_{i:g_i=0}, \tilde
{L}_{i:g_i=1})$'s contribution to the likelihood is
\begin{align*}
[p(n_i|L_i)p(L_i)]^{1-g_i}[p(\tilde{n}_i|\tilde{L}_i)p(\tilde
{L}_i)]^{g_i} & = \left[N(n_i; \eta_i, \delta^2)N(L_i; \nu_i, \psi
^2)\right]^{1-g_i} \\
&\times\left[\left\{1-\Phi\left(\frac{\tilde{n}_i -\eta_i}{\delta
}\right)\right\}\left\{1-\Phi\left(\frac{\tilde{L}_i - \nu_{i}}{\psi
}\right)\right\}\right]^{g_i}
\end{align*}
where $N(\cdot; m, s^2)$ denotes a Gaussian density function with mean
$m$ and variance $s^2$ and $\Phi(\cdot)$ denotes a standard normal cdf.
As a result, imputing $n_i$ for active players is carried out by first
imputing $L_i$ using a quadratic model with $d_i$.

We briefly note that although a Poisson model for $n_i$ might seem
natural, it is not appropriate in the current context as the
simultaneous increasing of the mean and variance of the Poisson
distribution seems to contradict what is empirically observed. Thus,
for simplicity, we elected to employ a Gaussian to model $n_i$ and
round the predictions (something that is not uncommon, see page 458 of
\citealt{BDA3}). Also, modeling 
$n_i$ non-parametrically could potentially improve prediction but we
elected to employ the simpler parametric model as its predictions were
satisfactory for our purposes. Nonetheless, predicting $n_i$ is of
considerable interest in its own right to NBA decision makers and could
be an interesting future research project.

Finally, given $n_i$ and letting $f_i(z_{it})$ denote the $i$th
player's underlying production curve value for the $t$th game played
(denoted by $z_{it}$), we model measurements $y_{it}$ as
\begin{align} \label{ymodel}
y_{it} = \beta_{0i} + f_{i}(z_{it}) + \epsilon_{it} \ \mbox{for} \ t =
1, \ldots, n_i \ (\tilde{n}_i \ \mbox{for} \ g_1 = 1)
\end{align}
where $ \epsilon_{it} \sim N(0,\sigma^2_i)$ independently across $i$.
It is possible that incorporating a more sophisticated error model
(such as autoregressive errors) could prove to be beneficial, but for
simplicity we maintain independence. A fairly popular method of
characterizing $f_i(\cdot)$ is to define a collection of basis
functions (e.g., wavelet, polynomial) and assume that $y_{it}$ lies in
their span. We adopt this method and employ a B-spline basis as it has
a number of attractive computational properties and facilitates active
player prediction as will be detailed shortly. Therefore, $f_i(\cdot)$
can be written as the following linear combination
\begin{align} \label{fmodel}
f_i(z_{it}) = \sum_{\ell=1}^{P_i} \beta_{i\ell} h_{\ell}(z_{it}; \bm{\xi
}_i) 
\end{align}
where $h_{\ell}(z;\xi_i)$ denotes the $\ell$th B-spline basis function
evaluated at knots contained in $\bm{\xi}_i$. If $p_i$ denotes the
number of inner knots and $q$ the spline degree, then $P_i = p_i + q +
1$. Now define $\bm{H}_i$ as the $n_i \times P_i$ matrix with rows $\{
h_1(z_{it}), \ldots, h_{P_i}(z_{it}) \}$ for $t = 1, \ldots, n_i \
(\tilde{n}_i \ \mbox{for} \ g_i=1)$, and $\bm{\beta}_i = \{\beta_{i1},
\ldots, \beta_{iP_i}\}$. Combining \eqref{ymodel} and \eqref{fmodel} produces
\begin{align}\label{LinearModel}
\bm{y}_i = \beta_{0i}\bm{1}_i + \bm{H}_i \bm{\beta}_i + \bm{\epsilon
}_{i} \ \mbox{for} \ \bm{\epsilon}_i \sim N_{}(\bm{0}, \sigma^2_i \bm{I}_{n_i}),
\end{align}
where $\bm{1}_i$ denotes a vector of ones and $\bm{I}_{n_i}$ an
identity matrix.

The dimension of $\bm{H}_i$ depends on $n_i$ (and $\tilde{n}_i$ for
active players). This coupled with the fact that B-splines form a local
basis in that each basis function is non-negative only on an interval
formed by $q + 2$ adjacent knots can be exploited to carry out active
player prediction. Since for any fixed $z_{it}$ at most $q+1$ basis
functions are positive, the predicted value of ${n}_i$ for active
players will determine the number of zero columns in $\bm{H}_i$. Thus,
the section of an active players curve corresponding to the $z_{it}$
values between $n_i$ and $\tilde{n}_i$ are completely informed by the
cluster specific curve or in the case that the player belongs to a
singleton, the grand mean curve (more details are in Section 3.3).
Using $\tilde{\bm{H}}_i$ to denote the design matrix that incorporates
the predicted value of $n_i$ based on $\tilde{n}_i$, the full
likelihood for $\bm{\Theta} = (\bm{\beta}, \bm{\beta}_0, \bm{\sigma}^2,
\bm{\eta},\bm{\nu}, \psi^2, \delta^2)$ is
\begin{align}
& \ell(\bm{y}_1, \ldots, \bm{y}_m,\bm{n},\bm{L}, \tilde{\bm{n}}, \tilde
{\bm{L}}, \bm{g}|\bm{\Theta}) \nonumber\\
&= \prod_{i=1}^n \left[N_{n_i}(\bm{y}_i; \beta_{0i}\bm{1}_i + \bm
{H}_i\bm{\beta}_i, \sigma^2_i \bm{I}_{n_i}) N(n_i; \eta_i, \delta
^2)N(L_i; \nu_i, \delta^2)\right]^{1-g_i} \times\nonumber\\
&\times \left[N_{\tilde{n}_i}(\bm{y}_i; \beta_{0i}\bm{1}_i + \tilde{\bm
{H}}_i\bm{\beta}_i, \sigma^2_i \bm{I}_{\tilde{n}_i}) \left\{1-\Phi\left
(\frac{\tilde{n}_i -\eta_i}{\delta}\right)\right\}\left\{1-\Phi\left
(\frac{\tilde{L}_i - \nu_{i}}{\psi}\right)\right\}\right]^{g_i}.
\end{align}

\subsection{Hierarchical Model}

The number and location of the inner-knots that make up $\bm{\xi}_i$
are rarely known. Their selection is crucial to producing an attractive
curve without over-fitting. So called free-knot splines is a very
flexible method that treats $\bm{\xi}_i$ as an unknown and has proved
to be quite parsimonious in knot selection. (\citealt
{BayesFreeKnotCurveFit} and \citealt{BayesNonlinearClassReg} provide a
nice overview.) Therefore, a possible direction to incorporating shape
variability in prediction as desired would be to base clustering on the
number and location of knots. However, to fit a free-knot spline some
type of transdimensional Markov Chain Monte Carlo (MCMC) algorithm is
often employed and this coupled with the PPMx prior for $\rho$ would
result in a doubly transdimensional MCMC algorithm that would become
prohibitively expensive. To avoid these computational issues and to
make the methodology more readily accessible, for each subject we
instead select a moderate number of equally spaced knots within the
knot domain and employ the Bayesian P-spline technology of \cite
{BayesianPsplines}. Now shape variability can influence clustering
through the penalty parameter of the P-splines. However, to retain
flexible subject-specific fits, we use P-splines as a prior
distribution of process level parameters and allow subject-specific
coefficients to vary around a cluster-specific mean. That is, we assume
the following process level structure for the $\bm{\beta}$'s:
\begin{align} \label{beta}
\bm{\beta}_i | \bm{\theta}^*_{s_i}, \lambda^{2*}_{s_i} & \sim N(\bm
{\theta}^*_{s_i}, \lambda^{2*}_{s_i}\bm{I}) \ \mbox{with} \ \sqrt
{\lambda^{2*}_{j}} \sim UN(0, A),
\end{align}
and use a Bayesian P-spline prior for the $\bm{\theta}^*_j$'s (with
$UN(\cdot, \cdot)$ denoting a Uniform distribution). A particularly
nice feature of the methodology is the explicit ability to control the
similarity between individual curves and their group counterparts
through the hyper-parameter $A$.

In order to highlight two departures from the Bayesian P-splines of
\citet{BayesianPsplines} required by the present modeling we very
briefly introduce them here. For more details see \citet
{BayesianPsplines} and \citet{BayesianSmoothing}. Bayesian P-splines
are the Bayesian analogue to splines penalized by $d$-order differences
and are constructed around $d$-order Gaussian random walks. For
example, for $d=1$
\begin{align}\label{rw}
\begin{split}
\theta^*_{j \ell} & = \theta^*_{j, \ell-1} + u_{j \ell} \ \ \ell= 2,
\ldots, n \\
\end{split}
\end{align}
with $u_{j\ell} \sim N(0, \tau^{2*}_j)$. Typically $p(\theta^*_{j1})
\propto v$, but an improper prior is not an appropriate probability
model for the Polya urn representation used in the PPMx. Thus, similar
to what was done in \cite{BayesianHierarchicalCurveRegistration} we assume
$\theta^*_{j1} \sim N(0, \tau^{2*}_j/v^2)$ (with analogous extensions
for $d>1$). The value $v$ can be assigned a prior distribution or be
set to a fixed value. Equation \eqref{rw} together with the $\theta
^*_{j1} \sim N(0, \tau^{2*}_j/v^2)$ produce $\bm{\theta}^*_j \sim N(\bm
{0}, \tau^{2*}_j \bm{K}^{-1})$ where $\bm{K}$ is a banded penalty
matrix with $v$ incorporated. $\tau^{2*}_j$ is the smoothing parameter
associated with Bayesian P-splines and is crucial in being able to
distinguish between individuals based on the smoothness of their
respective curves. As suggested by \cite{BayesianPsplines} we adopt
$\tau^{2*}_j \sim IG(a_{\tau},b_{\tau})$ where $IG(\cdot, \cdot)$
denotes an inverse Gamma distribution and $a_{\tau}$ and $b_{\tau}$ are
user supplied.

Recall that active player prediction is carried out by borrowing
strength among players in a cluster. If player $i$ belongs to a
singleton or all members of his cluster are active players, then at
least part of his prediction is completely guided by the prior on $\bm
{\theta}^*_j$. Since the prior is centered at $\bm{0}$ this would
produce poor active player predictions. To improve prediction in these
situations, we introduce $\bm{\mu}$ as a vector of global curve
coefficients such that $\bm{\theta}^*_j \sim N(\bm{\mu}, \tau^{2*}_j \bm
{K}^{-1})$ with $\bm{\mu} \sim N(\bm{0}, s^2_{\mu}\bm{I})$. Including
$\bm{\mu}$ potentially influences the values of $\bm{\theta}^*_j$ in
that smaller magnitudes achieve the same amount of smoothing as when
$\bm{\mu}=\bm{0}$. This should be taken into account when selecting
values for $a_{\tau}$ and $b_{\tau}$. Also, apart from improving
prediction, $\bm{\mu}$ is of interest in its own right as it provides
information regarding an average career curve among all
players.\looseness=1

We end the description of our Bayesian P-spline approach with details
regarding knot selection. A complicating factor of knot selection in
modeling these data is the massive misalignment associated with the
number of games played for each of the players. Making things worse is
the inherent discontinuities in games played through out the course of
one's career (e.g., the offseason, injuries, etc.) that we are not
considering. There is a ``curve registration'' literature dedicated to
better aligning functional domains in multi-subject studies (\citealt
{BayesianHierarchicalCurveRegistration}). However, we align career
paths by matching the percentile number of career games played. This is
carried out by transforming ``time" to the unit interval which greatly
simplifies the process of selecting $\bm{\xi}_i$. Therefore $z^*_{it} =
z_{it}/n_i$ is used instead of $z_{it}$. (For retired players $n_i$ is
the observed number of games played and the predicted for active
players.) Thus for retired players $z^*_{in_i} = 1$ while for active
players $z^*_{in_i} < 1$. Now $\bm{\xi}_i$ can be a knot set that
partitions the unit interval into equal subintervals and since it does
not depend on $n_i$ it can be the same for all players. We do note that
aligning career paths in this way is imperfect as the 95th percentile
of games played for one player might be during his third season while
for another player during his fifteenth season. Even so, we believe
that matching curves by way of percentile of games played produces
coherent comparisons and valid borrowing of strength. With an enriched
data set we could attempt to take into account possible discontinuities
in career paths. (This would actually be very interesting as many
players improve during the off season.) It would be fairly
straightforward to expand the model in a variety of ways, but the base
model proposed would continue being the work horse even as other more
idiosyncratic aspects of the data are considered.\looseness=1

With regards to modeling $\beta_{0i}$ and $\sigma^2_i$ there are any
number of ways one might proceed. It seems plausible that $\sigma^2_i$
might depend on $z_{it}$. That said, for sake of simplicity we utilize
the common prior structure for variance components $\sigma^2_i \sim
IG(a_{\sigma}, b_{\sigma})$ with $a_{\sigma}$ and $b_{\sigma}$ being
user supplied. For the subject-specific random intercepts, we use a
Gaussian-inverse-Gamma hierarchy such that $\beta_{0i} \sim N(\mu
_{b_0}, \sigma^2_{b_0})$ with $\mu_{b_0} \sim N(0, s^2_{b_0})$ and
$\sigma^2_{b_0} \sim IG(a_{b_0}, b_{b_0})$. Finally, typical conjugate
priors are used for $\bm{\alpha}=(\alpha_0, \alpha_1) \sim N(\bm{m}_a,
s^2_a\bm{I})$, $\bm{\gamma}=(\gamma_0, \gamma_1, \gamma_2) \sim N(\bm
{m}_{\gamma}, s^2_{\gamma}\bm{I})$, $\delta^2\sim IG(a_{\delta},
b_{\delta})$, and $\psi^2 \sim IG(a_{\psi}, b_{\psi})$.

Equation (\ref{HPPMx}) is provided to aid in visualizing how all the
moving parts of the hierarchical model are connected. Through out the
remainder of the paper we will refer to the entire hierarchical model
as HPPMx.\vadjust{\eject}
\begin{align}\label{HPPMx}
\bm{y}_i, n_i, L_i, \tilde{n}_i, \tilde{L}_i &| g_i, \bm{\beta}_i,\rho
,\sigma^2_i, \beta_{0i}, \bm{\alpha}, \bm{\gamma}, \delta^2,\psi^2
\nonumber\\
& \sim [N_{n_i}(\bm{y}_i; \beta_{0i}\bm{1}_i + \bm{H}_i\bm{\beta}_i,
\sigma^2_i \bm{I}_{n_i}) N(n_i; \eta_i, \delta^2)N(L_i; \nu_i, \psi
^2)]^{1 -g_i} \nonumber\\
& \times \Biggl[N_{\tilde{n}_i}(\bm{y}_i; \beta_{0i}\bm{1}_i + \tilde
{\bm{H}}_i\bm{\beta}_i, \sigma^2_i \bm{I}_{\tilde{n}_i})\notag\\
&\times \left\{1-\Phi
\left(\frac{\tilde{n}_i -\eta_i}{\delta}\right)\right\}\left\{1-\Phi
\left(\frac{\tilde{L}_i - \nu_{i}}{\psi}\right)\right\}\Biggr]^{g_i}
\nonumber\\
\sigma^2_i |a_{\sigma}, b_{\sigma} & \sim IG(a_{\sigma}, b_{\sigma})
\nonumber\\
\beta_{0i} | \mu_{b_0}, \sigma^2_{b_0} & \sim N(\mu_{b_0}, \sigma
^2_{b_0}) \ \ \mbox{with} \ \ \mu_{b_0} \sim N(0,s^2_{b_0}) \ \ \mbox
{and} \ \ \sigma^2_{b_0} \sim IG(a_{b_0}, b_{b_0}) \nonumber\\
\bm{\alpha} & \sim N(\bm{m}_a, s^2_a\bm{I}) \ \ \mbox{and} \ \ \delta^2
\sim IG(a_{\delta}, b_{\delta})\nonumber\\
\bm{\gamma} & \sim N(\bm{m}_{\gamma}, s^2_{\gamma}\bm{I}) \ \ \mbox
{and} \ \ \psi^2 \sim IG(a_{\psi}, b_{\psi}) \\
\bm{\beta}_i | \rho, \theta^*_{s_i}, \lambda^*_{s_i} & \sim N(\bm{\beta
}_i; \bm{\theta}^*_{s_i}, \lambda^{2*}_{s_i}) \ \ \mbox{with} \ \ \sqrt
{\lambda^{2*}_{h}} \sim UN(0, A) \nonumber\\
\bm{\theta}^*_h | \rho,\bm{\mu}, \tau_h^{2*}, \bm{K} & \sim N(\bm{\mu},
\tau^{2*}_h \bm{K}^{-1}) \ \ \mbox{with} \ \ \tau^{2*}_h \sim IG(a_{\tau
}, b_{\tau})\nonumber\\
\bm{\mu} & \sim N(\bm{0}, s^2_{\mu}\bm{I}) \nonumber\\
Pr(\rho) & \propto\prod_{h=1}^{k_m} c(S_j)g(\bm{x}^{\star}_j),\nonumber
\end{align}
for $i = 1, \ldots, m$ and $h = 1, \ldots, k_m$.

Before proceeding we make a brief comment regarding some specific model
components. Since the Bayesian P-splines are used at the prior level
rather than the process level of the hierarchical model, individual
curves are not directly influenced by its smoothing penalization. The
wiggliness of individual curves is a function of both $\tau^{2*}_j$ and
$A$. As $A$ increases the influence that $\tau^{2*}_j$ has on
individual curves decreases. This is investigated further in the
simulation study of Section 5. Therefore, if smooth individual curves
are desired together with large within group variability it may be
necessary to use 10-15 knots instead of the 20-30 knots recommended by
\citet{BayesianPsplines}.

\label{mcmc}
\section{Posterior Computation}
\subsection{MCMC Implementation}
We fit the proposed model to data by simulating a Markov chain whose
equilibrium distribution is the desired posterior distribution. The
algorithm employed is similar to \citet{MCMCSamplingMethodsForDPmixtureModels}'s
algorithm number 8 in that it
can be divided into two basic pieces. The first updates the partition
$\rho$ via the Polya urn scheme of \cite{BlackwellMacQueen} (and
further developed by \citealt{APredViewofBayesCluster}) and the other
updates the hierarchical model parameters using a Gibbs sampler
(\citealt{GemanGeman} and \citealt{GelfandSmith}) and Metropolis steps
(\citealt{Metropolis}).

To update the cluster membership for subject $i$, cluster weights are
created by comparing the unnormalized posterior for the $h$th cluster
when subject $i$ is excluded to that when subject $i$ is included. In
addition to weights for existing clusters, algorithm 8 of \citet
{MCMCSamplingMethodsForDPmixtureModels} requires calculating weights
for $p$ empty clusters whose cluster specific parameters are auxiliary
variables generated from the prior. To make this more concrete, let
$S_h^{-i}$ denote the $h$th cluster and $k^{-i}_m$ the number of
clusters when subject $i$ is not considered. Similarly $\bm{x}_h^{\star
-i}$ will denote the vector of covariates corresponding to cluster $h$
when subject $i$ has been removed. Then the multinomial weights
associated with the $k^{-i}_m$ existing clusters and the $p$ empty
clusters are
\begin{align*}
Pr(s_i = h | - ) \propto
\begin{cases}
N(\bm{\beta}_i ; \theta^{*}_{h}, \lambda^{2*}_{h}\bm{I}) \frac
{c(S_{h}^{-i}\cup\{i\})g(\bm{x}^{\star-i}_{h}\cup\{\bm{x}_i\}
)}{c(S_{h}^{-i})g(\bm{x}^{\star-i}_{h})}\ \mbox{for} \ h = 1, \ldots,
k^{-i}_m\\
N(\bm{\beta}_i; \bm{\theta}_{new,h}, \lambda^{2}_{new,h}\bm{I}) c(\{i\}
)g(\{\bm{x}_i\}) p^{-1}\ \mbox{for} \ h = k^{-i}_m+1, \ldots, k^{-i}_m
+ p.
\end{cases}
\end{align*}
Values for $\bm{\theta}_{new,h}$, $\lambda^2_{new,h}$ are auxiliary
variables drawn from their respective priors as required by algorithm
8. Care must be taken when subject $i$ belongs to a singleton cluster
as removing the $i$th subject produces an empty cluster. This in turn
requires relabeling the existing cluster specific components to avoid
gaps in the cluster labeling.

The full conditional distributions of $(\bm{\beta}_i, \bm{\beta}_0,
\sigma^2_i, \theta^*_j, \tau^{2*}_j)$ are fairly common derivations and
are provided in the Appendix. To update $(\lambda^*_j, \bm{\gamma}, \bm
{\alpha}, \delta^2, \psi^2)$ we employed a random walk Metropolis step
with a Gaussian proposal distribution. A Markov chain can be
constructed by employing a Gibbs sampler that first updates $\rho$ and
then on an individual basis updates model parameters by cycling through
each full conditional and using a Metropolis step for the non conjugate
parameters.

\subsection{Posterior Prediction Distributions} \label{PPD}
A particularly nice feature of using PPMx is the availability of career
prediction through covariate dependent predictive distributions. Using
PPMx, posterior predictive distributions are readily available and can
be obtained online in the sense that draws from this distribution can
be collected within the MCMC algorithm. The posterior predictive
distributions depend on covariate values through the allocation of a
new individual to one of the $k_m$ existing clusters or to a new
cluster using the following multinomial weights
\begin{align}
Pr(s_{n+1} = h | - ) \propto
\begin{cases}
\frac{c(S_{h}\cup\{n+1\})g(\bm{x}^{\star}_{h}\cup\{\bm{x}_{n+1}\}
)}{c(S_{h})g(\bm{x}^{\star}_{h})}\ \mbox{for} \ h = 1, \ldots, k_m\\
c(\{n+1\})g(\{\bm{x}_{n+1}\}) \ \mbox{for} \  h = k_m+1.
\end{cases}
\end{align}
Once the future player has been allocated to a cluster, one carries out
the typical Monte Carlo integration to sample from the posterior predictive.

\subsection[Predicting $n_i$]{Predicting $\bm{n}_i$}
Predicted values of ($n_i, L_i$) for retired players are produced at
each MCMC iteration. We essentially employ the multiple imputation
ideas of \cite{MissDat} but with the exception that we are very much
interested in the values being imputed. Predictions are fairly straight
forward as they only depend on the full conditionals of $n_i$ and $L_i$
which turn out to be truncated normal with $\tilde{n}_i$ and $\tilde
{L}_i$ acting as lower bounds (the full conditionals are provided in
the Appendix).

\section{Simulation Studies}
We conduct two small simulation studies to investigate the behavior of
the HPPMx model (\ref{HPPMx}). Recall that the principal motivation in
incorporating a hierarchy is to balance goodness of individual curve
fits with the production of meaningful clusters which facilitate
prediction. Therefore apart from showing improved prediction
performance, the simulation study explores just how much goodness of
individual fit is sacrificed in the name of prediction (which is very
little as will be seen). The first simulation study demonstrates the
method's superior predictive performance by comparing out of sample
mean integrated prediction rates to that of two competitors (which are
detailed shortly). The second explores how certain model components
influence subject-specific fits, curve smoothness, and clustering. The
two competitors selected represent the extremes HPPMx attempts to
balance, namely fitting each player independently versus assigning
players cluster-specific curves. The first competitor is a
semi-parametric regression model (henceforth SP) that fits individual
curves independently and flexibly. The second is a semi-parametric
regression model with a Dirichlet process prior (henceforth SPDP) which
produces individual curves that are cluster specific. More precisely we consider
\begin{align}\label{competitor}
y_{it} & = \bm{x}'_i \bm{\beta} + f_i(z_{it} ) + \epsilon_{it} \ \mbox
{with} \ \epsilon_{it} \sim N(0, \sigma^2_i) \ \mbox{for} \ i = 1,\ldots
, m \ \mbox{and} \ t=1,\ldots,n
\end{align}
where $(f_i(z_{i1}), \ldots, f_i(z_{in}))' = \bm{H}\bm{\theta}_i$ is
modeled using subject-specific linear combinations of B-spline basis
functions, $z_{it} \in[0,1]$, $\bm{x}_i$ is a vector of covariates
that will be described shortly and
\begin{align*}
& \mbox{\underline{SP}} & & \mbox{\underline{SPDP}} \\
\bm{\theta}_i & \sim N(\bm{0}, \tau^2\bm{K}^{-1}) & \bm{\theta}_i | G &
\sim G \\
& & G & \sim DP(M, G_0) \ \mbox{with} \ G_0 = N(\bm{0}, \tau^2\bm{K}^{-1}).
\end{align*}
Notice that under SP a P-spline prior is used for $\bm{\theta}_i$ while
under SPDP the base measure of the DP is a P-spline prior. As with
HPPMx, $\tau^2 \sim IG(a_{\tau}, b_{\tau})$. The competitors selected,
though reasonable, aren't capable of providing active player
predictions. Therefore, prediction assessment in the simulation study
is only carried out for career prediction. Finally, we investigate the
influence that covariates have on clustering by considering the HPPMx
model with a PPM prior rather than a PPMx prior (which will be hence
forth referred to as HPPM).

%
\begin{figure}[htbp]
\includegraphics{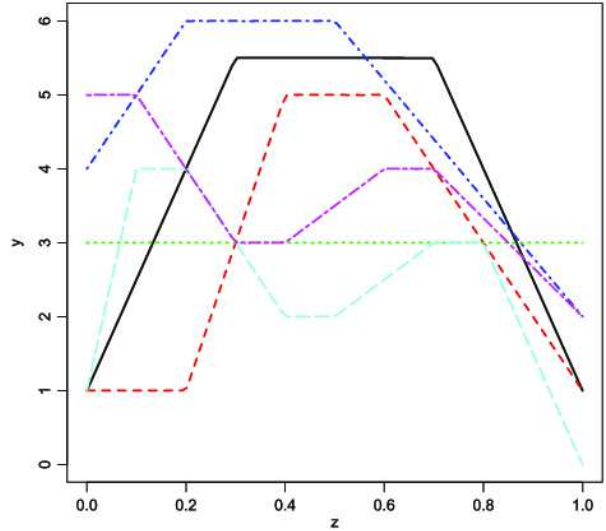}
\caption{The six mean curves used in the simulation study.}
\label{SimFunc}
\end{figure}

Since both simulation studies employ the same general data generating
mechanism we provide details here. A response vector is generated using
\begin{align}\label{DGM}
y_{it} = b_{0i} + f_{group_i}(z_{it}) + \epsilon_{it} \ \mbox{with} \
\epsilon_{it} \sim N(0, w^2)
\end{align}
where $f_{group_i}(\cdot)$ corresponds to $group_i = 1, \ldots, 6$
possible mean curves which were created using the NBA data as a guide
(see Figure \ref{SimFunc}). The six mean curves are made to depend on
covariates by creating two categorical covariates that when crossed
produce six categories, one for each mean curve. A continuous covariate
was generated by $x^*_i \sim N(group_i, 0.1)$. Since $x^*_i$ depends on
the two categorical covariates, an interaction between them is created.
The three covariates were included in all model fits. Lastly, the
random intercept is generated using $b_{0i} \sim N(10,2)$.

The factors we explore in the simulation study with their respective
levels are%
\begin{itemize}
\item value for hyper parameter $A$ (0.1, 1, 10)
\item number of knots (5,15,30)
\item variance of (\ref{DGM}) ($w^2 = 0.1$, $w^2=1$)
\item number of observations per subject ($n=50$, $n=100$).
\end{itemize}
$A$ and the number of knots were selected to investigate how P-splines
function as a prior at the process level instead of at the observation
level of a hierarchical model. With $n$ and $w^2$ we see how the
methodology performs as more information becomes available relative to noise.
For each combination of the factor levels 100 data sets with $m=60$ (10
players per group) are generated and for each data set SP, SPDP, HPPM,
and HPPMx are fit.

For all four procedures we set $a_{\tau} = 1$, $b_{\tau} = 0.05$,
$a_{\sigma} = b_{\sigma} = 1.0$ and $v=1$. For PPMx and PPM $s^2_{b_0}
= s^2_{\mu} = 100^2$, and for SP and SPDP $\bm{\beta} \sim N(\bm{0},
100^2\bm{I})$. Finally for HPPMx, the cohesion and similarity functions
employed are those that match the marginal prior on partitions implied
by a DP prior (see Section 6.1 for more details). Each of the four
procedures were fit to each synthetic data set using 1000 MCMC iterates
after discarding the first 5000 as burn-in. Empirically based starting
values were employed which accelerated convergence making the 5000
iterate burn-in sufficient.

\subsection{Out of Sample (Career) Prediction}
To assess out of sample (career) prediction, for each of the 100
generated data sets 100 additional out of sample subjects were
generated. The $f_{group}(\cdot)$ associated with each new subject is
known and therefore out of sample prediction can be assessed by
comparing $\hat{f}(\cdot)$ from the four procedures to $f_{group}(\cdot
)$. After centering both $f_{group_j}(\cdot)$ and $\hat{f}_j(\cdot)$
(i.e. subtract off the empirical mean) for the $j$th out of sample
subject, we measure prediction accuracy using the mean integrated
squared prediction error
\begin{align}\label{IMSE}
MISPE_j = E \int[\hat{f}_j(z) - f_{group_j}(z)]^2dz \approx \sum
_{t}\Delta_tE[\hat{f}(z_{jt}) - f_{group}(z_{jt})]^2
\end{align}
where $\Delta_t = (z_{jt+1} - z_{jt})$. Equation (\ref{IMSE})
essentially measures the average squared area between $f_{group_j}(\cdot
)$ and $\hat{f}_j(\cdot)$ for the $j$th out of sample player over $z$'s
domain. The values in Table \ref{OSP} correspond to
\begin{align}\label{mnIMSE}
\frac{1}{D}\sum_{d=1}^D\frac{1}{100}\sum_{j=1}^{100} MISPE_{dj}
\end{align}
where $d$ indexes the $D=100$ generated data sets.

%
\begin{table}[t]
\caption{Results from the simulation study investigating out of sample
prediction. Table entries are calculated using (\ref{mnIMSE}) with $m =
60$ players.}\label{OSP}
\vspace*{4pt}
\begin{tabular}{l l l cccccc}\hline
&&&\multicolumn{3}{c}{$n=50$} & \multicolumn{3}{c}{$n=100$}\\ \cline
{4-6} \cline{7-9}
&&&\multicolumn{3}{c}{Number of knots} & \multicolumn{3}{c}{Number of
knots}\\ \cline{4-6} \cline{7-9}
\multicolumn{2}{c}{} & Model & 5 & 15 & 30 & 5 & 15 & 30\\\hline
\multirow{12}{*}{$w^2 = 0.1$} &\multirow{4}{*}{$A=0.1$} &HPPMx & 0.131
& 0.119 & 0.123 & 0.149 & 0.144 & 0.122 \\
& & HPPM & 0.891 & 0.868 & 0.847 & 0.873 & 0.855 & 0.848 \\
& & SP & 1.310 & 1.301 & 1.319 & 1.297 & 1.288 & 1.277 \\
& & SPDP & 0.782 & 0.784 & 0.789 & 0.780 & 0.779 & 0.775 \\ \cline{2-9}
&\multirow{4}{*}{$A=1$} & HPPMx & 0.034 & 0.026 & 0.095 & 0.026 & 0.025
& 0.029 \\
& & HPPM & 0.791 & 0.795 & 0.799 & 0.779 & 0.784 & 0.777 \\
& & SP & 1.324 & 1.319 & 1.306 & 1.285 & 1.300 & 1.279 \\
& & SPDP & 0.785 & 0.789 & 0.782 & 0.779 & 0.782 & 0.776 \\ \cline{2-9}
&\multirow{4}{*}{$A=10$} & HPPMx & 0.022 & 0.025 & 0.108 & 0.023 &
0.041 & 0.037 \\
& & HPPM & 0.786 & 0.791 & 0.800 & 0.774 & 0.776 & 0.784 \\
& & SP & 1.321 & 1.307 & 1.292 & 1.282 & 1.278 & 1.289 \\
& & SPDP & 0.785 & 0.783 & 0.779 & 0.775 & 0.777 & 0.780 \\ \hline
\multirow{12}{*}{$w^2 = 1$}&\multirow{4}{*}{$A=0.1$} &HPMMx & 0.155 &
0.203 & 0.254 & 0.142 & 0.151 & 0.242 \\
& & HPMM & 0.882 & 0.848 & 0.852 & 0.854 & 0.828 & 0.837 \\
& & SP & 1.315 & 1.312 & 1.326 & 1.277 & 1.287 & 1.283 \\
& & SPDP & 0.783 & 0.786 & 0.787 & 0.771 & 0.776 & 0.777 \\ \cline{2-9}
&\multirow{4}{*}{$A=1$} & HPPMx & 0.100 & 0.102 & 0.170 & 0.045 & 0.069
& 0.102 \\
& & HPPM & 0.796 & 0.807 & 0.829 & 0.778 & 0.789 & 0.795 \\
& & SP & 1.312 & 1.316 & 1.306 & 1.257 & 1.279 & 1.288 \\
& & SPDP & 0.783 & 0.785 & 0.785 & 0.766 & 0.776 & 0.776 \\ \cline{2-9}
&\multirow{4}{*}{$A=10$} & HPPMx & 0.080 & 0.095 & 0.157 & 0.047 &
0.066 & 0.113 \\
& & HPPM & 0.823 & 0.817 & 0.836 & 0.798 & 0.794 & 0.809 \\
& & SP & 1.331 & 1.305 & 1.327 & 1.291 & 1.269 & 1.284 \\
& & SPDP & 0.789 & 0.780 & 0.787 & 0.778 & 0.772 & 0.776 \\
\hline
\end{tabular}
\end{table}

From Table \ref{OSP} we see that HPPM and SPDP provide similar
predictions which is to be expected as both employ a DP prior (although
not at the same level of a hierarchy). What should be very obvious is
that HPPMx does a much better job in out of sample prediction relative
to the other three procedures for all data generating scenarios.

\subsection{Goodness of Individual Fits, Curve Smoothness, and Clustering}
To assess goodness of individual fits we employ the following $R^2$
type goodness-of-fit statistic from \cite{FunctionalDataAnalysisBook}:
\begin{align}\label{R2}
R_i^2 = 1 - \frac{\sum_t(\hat{f}_i(z_{it}) - y_{it})^2}{\sum
_t(y_{it}-\bar{y}_{i} )^2}.
\end{align}
$R^2_i$ can be loosely interpreted as a coefficient of determination in
that as $R^2_i$ approaches 1, individual fits improve. Negative values
of $R^2_i$ indicate that $\bar{y}_i$ predicts better than $\hat
{f}_i(\cdot)$. The values in Table \ref{SSr2} correspond to
\begin{align}\label{mnR2}
\frac{1}{D}\sum_{d=1}^D\frac{1}{m}\sum_{i=1}^m R_{di}^2.
\end{align}

%
\begin{table}[t]
\caption{Results from the simulation study investigating
goodness-of-fit. Table entries are calculated using (\ref{mnR2}) with
$m = 60$ players.}\label{SSr2}
\vspace*{4pt}
\begin{tabular}{l l l cccccc}
\hline
&&&\multicolumn{3}{c}{$n=50$} & \multicolumn{3}{c}{$n=100$}\\ \cline
{4-6} \cline{7-9}
&&&\multicolumn{3}{c}{Number of knots} & \multicolumn{3}{c}{Number of
knots}\\ \cline{4-6} \cline{7-9}
\multicolumn{2}{c}{} & Model & 5 & 15 & 30 & 5 & 15 & 30\\
\hline
\multirow{12}{*}{$w^2 = 0.1$} &\multirow{4}{*}{$A=0.1$} &HPPMx & 0.979
& 0.990 & 0.950 & 0.981 & 0.989 & 0.984 \\
& & HPPM & 0.813 & 0.908 & 0.919 & 0.812 & 0.939 & 0.952 \\
& & SP & 0.985 & 0.993 & 0.994 & 0.984 & 0.992 & 0.992 \\
& & SPDP & 0.802 & 0.794 & 0.774 & 0.764 & 0.759 & 0.743 \\ \cline{2-9}
&\multirow{4}{*}{$A=1$} & HPPMx & 0.981 & 0.965 & 0.864 & 0.984 & 0.991
& 0.979 \\
& & HPPM & 0.983 & 0.987 & 0.950 & 0.984 & 0.991 & 0.989 \\
& & SP & 0.985 & 0.993 & 0.994 & 0.984 & 0.992 & 0.992 \\
& & SPDP & 0.808 & 0.811 & 0.798 & 0.771 & 0.743 & 0.746 \\ \cline{2-9}
&\multirow{4}{*}{$A=10$} & HPPMx & 0.984 & 0.975 & 0.853 & 0.984 &
0.990 & 0.979 \\
& & HPPM & 0.985 & 0.987 & 0.937 & 0.984 & 0.991 & 0.992 \\
& & SP & 0.985 & 0.993 & 0.994 & 0.984 & 0.992 & 0.992 \\
& & SPDP & 0.802 & 0.793 & 0.787 & 0.765 & 0.747 & 0.750 \\ \hline
\multirow{12}{*}{$w^2 = 1$}&\multirow{4}{*}{$A=0.1$} &HPPMx & 0.583 &
0.593 & 0.572 & 0.574 & 0.593 & 0.575 \\
& & HPPM & 0.485 & 0.537 & 0.497 & 0.476 & 0.552 & 0.541 \\
& & SP & 0.613 & 0.663 & 0.713 & 0.570 & 0.575 & 0.533 \\
& & SPDP & 0.538 & 0.554 & 0.565 & 0.515 & 0.521 & 0.520 \\ \cline{2-9}
&\multirow{4}{*}{$A=1$} & HPPMx & 0.585 & 0.608 & 0.630 & 0.572 & 0.589
& 0.605 \\
& & HPPM & 0.589 & 0.618 & 0.663 & 0.574 & 0.596 & 0.616 \\
& & SP & 0.613 & 0.662 & 0.711 & 0.566 & 0.575 & 0.531 \\
& & SPDP & 0.541 & 0.555 & 0.562 & 0.511 & 0.522 & 0.523 \\ \cline{2-9}
&\multirow{4}{*}{$A=10$} & HPPMx & 0.587 & 0.610 & 0.626 & 0.574 &
0.589 & 0.607 \\
& & HPPM & 0.593 & 0.624 & 0.667 & 0.577 & 0.597 & 0.617 \\
& & SP & 0.613 & 0.662 & 0.712 & 0.568 & 0.574 & 0.535 \\
& & SPDP & 0.539 & 0.554 & 0.561 & 0.514 & 0.519 & 0.522 \\
\hline
\end{tabular}
\end{table}

From Table \ref{SSr2} we see that SP tends to produce the best
individual fits and SPDP the worst. This is of course expected as all
individuals are fit independently by SP while SPDP provides cluster
specific curves. However, HPPMx does remarkably well in producing good
individual fits as HPPMx is very close to SP particularly as $n$
increases. Thus, HPPMx's meaningful cluster production doesn't require
sacrificing much goodness of individual fit.

%
\begin{table}[t]
\caption{Results from the simulation study investigating smoothness.
Table entries are calculated using (\ref{mnlSD}) with $m = 60$
players.}\label{SSsmooth}
\vspace*{4pt}
\begin{tabular}{l l l cccccc}
\hline
&&&\multicolumn{3}{c}{$n=50$} & \multicolumn{3}{c}{$n=100$}\\ \cline
{4-6} \cline{7-9}
&&&\multicolumn{3}{c}{Number of knots} & \multicolumn{3}{c}{Number of
knots}\\ \cline{4-6} \cline{7-9}
\multicolumn{2}{c}{} & Model & 5 & 15 & 30 & 5 & 15 & 30\\
\hline
\multirow{12}{*}{$w^2 = 0.1$} &\multirow{4}{*}{$A=0.1$} &HPPMx & 0.162
& 0.165 & 0.171 & 0.082 & 0.082 & 0.087 \\
& & HPPM & 0.143 & 0.154 & 0.165 & 0.073 & 0.078 & 0.084 \\
& & SP & 0.162 & 0.162 & 0.166 & 0.081 & 0.080 & 0.083 \\
& & SPDP & 0.154 & 0.153 & 0.157 & 0.077 & 0.076 & 0.079 \\ \cline{2-9}
&\multirow{4}{*}{$A=1$} & HPPMx & 0.163 & 0.165 & 0.170 & 0.081 & 0.081
& 0.084 \\
& & HPPM & 0.163 & 0.164 & 0.168 & 0.081 & 0.081 & 0.085 \\
& & SP & 0.162 & 0.162 & 0.166 & 0.081 & 0.080 & 0.083 \\
& & SPDP & 0.154 & 0.154 & 0.158 & 0.077 & 0.076 & 0.079 \\ \cline{2-9}
&\multirow{4}{*}{$A=10$} & HPPMx & 0.163 & 0.165 & 0.169 & 0.082 &
0.081 & 0.085 \\
& & HPPM & 0.162 & 0.164 & 0.168 & 0.082 & 0.081 & 0.086 \\
& & SP & 0.161 & 0.162 & 0.166 & 0.081 & 0.080 & 0.083 \\
& & SPDP & 0.153 & 0.154 & 0.158 & 0.077 & 0.076 & 0.079 \\ \hline
\multirow{12}{*}{$w^2 = 1$}&\multirow{4}{*}{$A=0.1$} &HPPMx & 0.179 &
0.208 & 0.274 & 0.096 & 0.107 & 0.132 \\
& & HPPM & 0.155 & 0.182 & 0.250 & 0.082 & 0.096 & 0.120 \\
& & SP & 0.195 & 0.206 & 0.258 & 0.088 & 0.071 & 0.062 \\
& & SPDP & 0.172 & 0.174 & 0.191 & 0.095 & 0.090 & 0.095 \\ \cline{2-9}
&\multirow{4}{*}{$A=1$} & HPPMx & 0.184 & 0.210 & 0.288 & 0.097 & 0.109
& 0.136 \\
& & HPPM & 0.183 & 0.209 & 0.298 & 0.097 & 0.108 & 0.140 \\
& & SP & 0.198 & 0.205 & 0.258 & 0.088 & 0.071 & 0.061 \\
& & SPDP & 0.174 & 0.175 & 0.190 & 0.094 & 0.090 & 0.095 \\ \cline{2-9}
&\multirow{4}{*}{$A=10$} & HPPMx & 0.184 & 0.212 & 0.286 & 0.097 &
0.110 & 0.137 \\
& & HPPM & 0.185 & 0.210 & 0.296 & 0.100 & 0.108 & 0.141 \\
& & SP & 0.197 & 0.206 & 0.258 & 0.088 & 0.071 & 0.062 \\
& & SPDP & 0.173 & 0.175 & 0.189 & 0.094 & 0.090 & 0.095 \\
\hline
\end{tabular}
\end{table}

To assess curve smoothness we calculate the standard deviation of the
lag one differences from the estimated curve
\begin{align}\label{lSD}
{\ell}SD_i = \sqrt{\frac{1}{n-3}\sum_{t=1}^{n-1} (lag_{it} - \overline{lag}_i)^2},
\end{align}
where $lag_{it} = \hat{f}_i(z_{it+1}) - \hat{f}_i(z_{it}) \ \mbox{for}
\  t = 1, \ldots, n-1$ and $\overline{lag}_i = 1/(n-1)\sum
_{t=1}^{n-1}lag_{it}$. Large values of ${\ell}SD_i$ generally indicate
more wiggliness relative to small values. Values provided in Table \ref
{SSsmooth} correspond to
\begin{align}\label{mnlSD}
\frac{1}{D}\sum_{d=1}^D\frac{1}{m}\sum_{i=1}^m {\ell} SD_{di}.
\end{align}

%
\begin{table}[t]
\caption{Results from the simulation study investigating cluster
estimation. Table entries correspond to the number of estimated
clusters averaged over 100 simulated data sets. }\label{SScluster}
\vspace*{4pt}
\begin{tabular}{l l l cccccc}
\hline
&&&\multicolumn{3}{c}{$n=50$} & \multicolumn{3}{c}{$n=100$}\\ \cline
{4-6} \cline{7-9}
&&&\multicolumn{3}{c}{Number of knots} & \multicolumn{3}{c}{Number of
knots}\\ \cline{4-6} \cline{7-9}
\multicolumn{2}{c}{} & Model & 5 & 15 & 30 & 5 & 15 & 30\\
\hline
\multirow{6}{*}{$w^2 = 0.1$} &\multirow{2}{*}{$A=0.1$} &HPPMx & 5.98 &
5.98 & 5.99 & 5.97 & 5.98 & 5.96 \\
& & HPPM & 3.93 & 4.26 & 4.49 & 4.32 & 4.37 & 4.40 \\ \cline{2-9}
&\multirow{2}{*}{$A=1$} & HPPMx & 9.24 & 8.53 & 6.52 & 10.93 & 10.26 &
9.02 \\
& & HPPM & 9.49 & 9.18 & 7.89 & 10.85 & 10.77 & 10.02 \\ \cline{2-9}
&\multirow{2}{*}{$A=10$} & HPPMx & 10.32 & 8.55 & 6.42 & 13.39 & 10.34&
8.86 \\
& & HPPM & 10.45 & 8.31 & 7.14 & 11.84 & 9.67 & 8.94 \\ \hline
\multirow{6}{*}{$w^2 = 1$}&\multirow{2}{*}{$A=0.1$} &HPPMx & 5.97 &
5.96 & 5.99 & 5.98 & 5.97 & 5.99 \\
& & HPPM & 4.09 & 4.36 & 4.72 & 4.02 & 4.54 & 4.66 \\ \cline{2-9}
&\multirow{2}{*}{$A=1$} & HPPMx & 7.25 & 6.87 & 6.11 & 8.20 & 7.94 &
6.82 \\
& & HPPM & 7.08 & 6.74 & 6.15 & 7.94 & 7.77 & 7.40 \\ \cline{2-9}
&\multirow{2}{*}{$A=10$} & HPPMx & 7.24 & 6.89 & 6.28 & 8.90 & 8.06 &
7.08 \\
& & HPPM & 5.84 & 6.03 & 5.36 & 7.14 & 6.88 & 6.94 \\
\hline
\end{tabular}
\end{table}

From Table \ref{SSsmooth} it appears that curves become less wiggly as
$n$ increases relative to the number of knots. This is expected. Also
unsurprising is that HPPMx and HPPM produce similar values of $(\ref
{mnlSD})$ with the biggest differences occurring when $w^2$ (within
player variability) and $A$ (within cluster variability) are small.
What is a bit surprising is that the value of $A$ doesn't much alter
curve smoothness for HPPMx. It appears that $w^2$ is more influential.
Overall, since the values of (\ref{mnlSD}) for HPPMx are fairly similar
to those for SP and SPDP (recall that SP and SPDP are not influenced by
$A$), penalizing curves directly with a P-spline prior produces curves
with similar smoothness as those produced through the hierarchical model.

To see how the PPMx prior improves clustering relative to the PPM
prior, Table \ref{SScluster} provides the number of estimated clusters
($k_m$) averaged over all $D=100$ data sets. For each data set $\rho$
was estimated using \cite{dahl:2006}'s method which is based on
least-squares distance from the matrix of posterior pairwise
co-clustering probabilities (note that an estimate of $\rho$ also
provides an estimate of $k_m$).

The true value of $k_m$ in Table \ref{SScluster} is six for all
scenarios. It appears that as $n$ increases, the PPMx prior tends to
converge to the six clusters faster than the PPM prior. It also appears
that the clustering mechanisms of the HPPMx and HPPM depend on $A$.
This is to be expected however, because as $A$ increases curves are
allowed to deviate further from cluster specific means, thus creating
more clusters.

\vspace*{-2pt}\section{Analysis and Results} \label{results}\vspace*{-1pt}
In this we section provide results of fitting HPPMx to the NBA data set.

\vspace*{-1pt}\subsection{Model Details and Prior Selection}\vspace*{-1pt}

We first provide a bit of detail regarding cohesion and similarity
functions used and then on prior values. The cohesion and similarity
functions employed match the PPMx prior\vadjust{\eject} to the marginal prior on
partitions implied by the DP prior. This results in an {a priori}
clustering of a few large clusters that represent typical player
production and a few smaller clusters of ``abnormal'' players. Thus, we
set $c(S_j) = M(|S_j| - 1)!$ with $M=1$ favoring a small number of
clusters. The similarity functions used are typical conjugate models
for continuous and categorical variables with parameter values
suggested by \cite{PPMxMullerQuintanaRosner} resulting in
\label{gfunc}
\begin{align*}
g(\bm{x}^{\star}_j) & = g_1(\bm{x}^{\star}_{j1})g_2(\bm{x}^{\star
}_{j2})g_3(\bm{x}^{\star}_{j3}) \\
& = \int\prod_{i\in S_j} N(x_{i1}; m_j, 1) N(m_j; 0, 10) \pi
_{i,x_{i2}} \\
& \times Dir(\pi_{i,x_{i2}}; 0.1, 0.1, 0.1) \pi_{i,x_{i3}}Dir(\pi
_{i,x_{i3}};0.1, 0.1,0.1)dm_jd\bm{\pi}_{1j}d\bm{\pi}_{2j} \\
& = \frac{N_{n_j}(\bm{x}^{\star}_j; \bm{0}, \bm{I}) N(0; 0, 10)}{N(\hat
{m}; 0, \hat{s}^2)}\frac{\Gamma(\sum_{c=1}^3 n_{1jc} + 0.1)}{\prod
_{c=1}^3\Gamma(n_{1jc} + 0.1)}\frac{\Gamma(\sum_{c=1}^3 n_{2jc} +
0.1)}{\prod_{c=1}^3\Gamma(n_{2jc} + 0.1)}.
\end{align*}
$\pi_{i,x_{i2}}$ and $\pi_{i,x_{i3}}$ denote $x_{i2}$ and $x_{i3}$'s
probability vector, $n_{1jc}$ are the number of players in cluster $j$
that have covariate value $c$ for $x_{i2}$ and $n_{2jc}$ the number of
players for $x_{i3}$ (as a reminder, $x_{i1}$ corresponds to age,
$x_{i2}$ experience and $x_{i3}$ draft order). In addition, $\hat{s}^2
= (n_j + 1/10)^{-1}$ and $\hat{m} = \hat{s}^2\bm{1}'\bm{x}^{\star}_{1j}$.

A first-order Bayesian P-spline prior was used and following
suggestions in \citet{BayesianPsplines}, we set $a = 1$ and $b = 0.05$.
We found that results were fairly robust to reasonable prior
specifications of $\tau^2_h$. From the simulation study setting $A = 1$
seemed reasonable so that individual curves are fairly similar to their
cluster-specific counterparts. Preliminary investigations indicated
that methodology is robust to variance prior specifications so with
hopes of being diffuse we set $a_{\sigma} = b_{\sigma} = a_{\delta} =
b_{\delta} = a_{\psi}=b_{\psi} = 1$ and $s^2_{b_0} = 100^2$. To produce
a flat prior for $\bm{\gamma}$ we use $\bm{m}_{\gamma} = \bm{0}$ and
$s^2_{\gamma} = 100^2$. Since there are 82 games in an NBA season we
set $m_{a_1} = 76$ (taking into account missed games due to injury)
with $s^2_{a} = 10^2$.

The MCMC algorithm was run until 1000 iterates from a Markov chain were
collected after discarding the first 25,000 as burn in and thinning by
25. Convergence was monitored using MCMC iterate trace plots.

\subsection{Fits of Individual Player Production Curves}
\begin{figure}[htbp]
\includegraphics{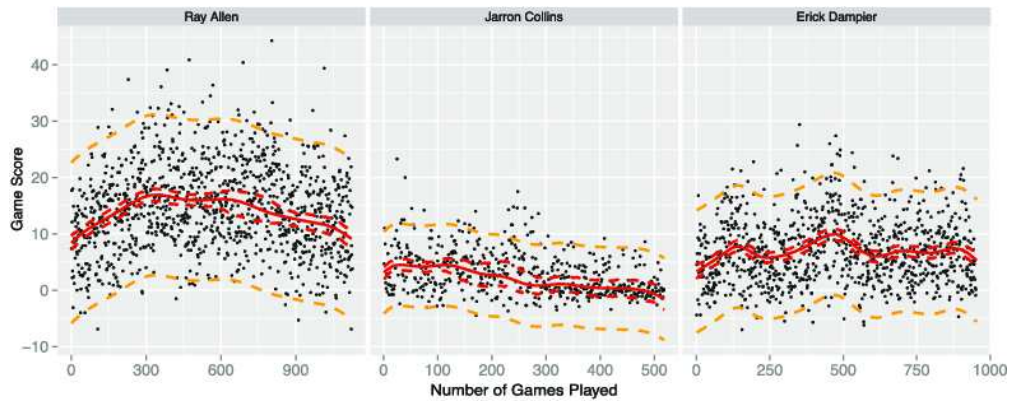}
\caption{Posterior fits for three NBA players. The solid red lines are
point-wise posterior mean curves, the dashed red lines are 95\% mean
point-wise credible intervals, and the dashed orange lines are 95\%
point-wise prediction intervals.}
\label{IndividualPosteriorFits}
\end{figure}

Figure \ref{IndividualPosteriorFits} displays the posterior mean curves
with 95\% credible and prediction bands for the three players
introduced in Figure \ref{rawscatterplot}. Notice that even though the
fits are fairly flexible they are smoothed relative to the loess fits
provided in Figure \ref{rawscatterplot}.

%

%
\begin{figure}[t!]
\includegraphics{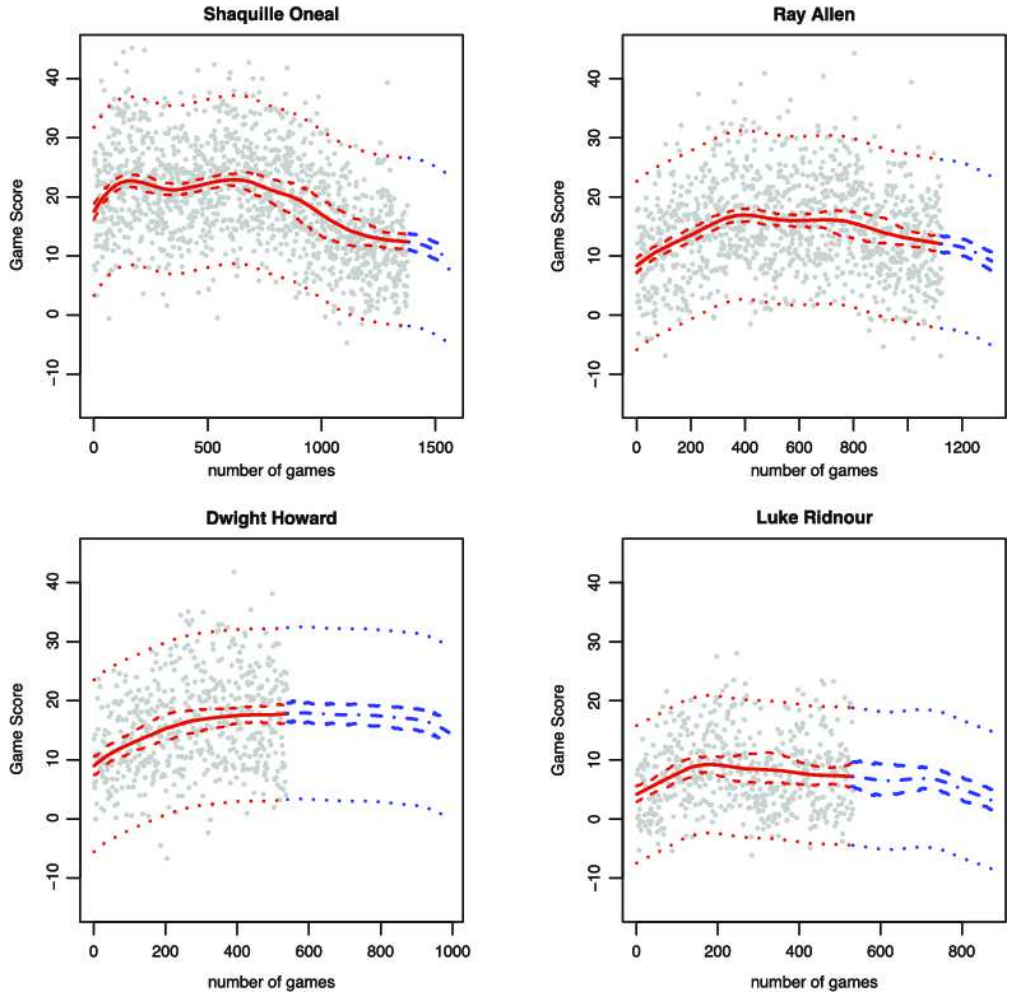}
\caption{Active player predictions for four NBA players. Lines
associated with active player prediction are blue. The dashed lines
represent point-wise 95\% mean credible bands, and the dotted lines
95\% point-wise prediction bands.}
\label{ACP}
\vspace*{-6pt}
\end{figure}

\subsection{Active Player Prediction}
Displaying the results of active player prediction in and of itself is
not trivial as curves depend completely on the predicted values of
$n_i$. To simplify the process we display the active player prediction
curves conditioned on $E(n_i|\bm{y}_i)$. This requires producing a
curve conditioned on $E(n_i|\bm{y}_i)$ for each MCMC iterate of $\bm
{\beta}$. From these MCMC iterates, we estimate an average prediction
curve with point-wise 95\% credible bands and prediction bands. Figure
\ref{ACP} contains the estimated mean curve with credible bands and
prediction bands corresponding to four players in varying stages of
their career. Shaquille O'Neal played beyond the 2009/2010 season but
has since retired. His average number of predicted games played turned
out to be 1545 and the actual number of games played is 1423 (including
playoffs). Ray Allen continues to play but is nearing the end of his
career and the predicted sharp decrease in production mirrors reality.
Dwight Howard and Luke Ridnour are two completely different types of
players and are provided as a means to demonstrate the flexibility in
the predictions. $E(n_i|\bm{y}_i)$ for Dwight Howard is quite
conservative and barring injury should under estimate his career game
total, while $E(n_i|\bm{y}_i)$ for Beno Udrih is quite reasonable.
Regardless, the four predictions display completely plausible decreases
in production as the players approach retirement.

%
\begin{figure}[htbp]
\includegraphics{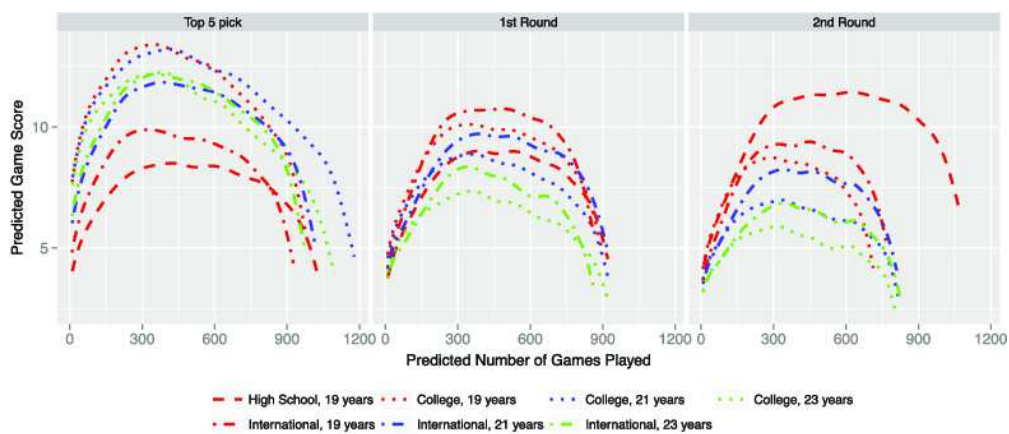}
\caption{Career prediction curves for different levels of draft order,
playing experience and age during first game played. }
\label{careerPrediction}
\end{figure}

\subsection{Career Prediction Analysis}
For career prediction we employ the predictive distributions as
described in Section \ref{PPD}\vadjust{\eject} resulting in the curves found in Figure
\ref{careerPrediction}. We include the High School level of
experience
even though the latest CBA requires at least one year post high school
experience before being drafted. For age during first game played, we
considered 19, 21, and 23 years old. (We do not consider ages 21 and 23
for High School level of experience as those scenarios are practically
impossible.) The curves are presented conditioned on the predicted
number of games played averaged over all active players that belong to
each respective group.

Before describing results it is important to keep in mind that only
players who played at least three years were included in the analysis.
This explains the seemingly high predictions for second round picks.
Also, from Table \ref{playercategorysummary} it can be seen that only
one player (considered in the analysis) was drafted in the second round
straight from high school (Rashard Lewis). So you will notice that the
predicted curve for this group follows a trajectory similar to that of
Rashard Lewis's career. Even with that being the case, a few
interesting trends emerge. It appears that there is much more
variability in curve location for Top Five Picks. Also the players that
are Top Five Picks tend to reach their max production earlier in their
career. Age clearly influences a player's production as players that
start their career at a younger age tend to have higher production
rates. It appears that the shapes of curves vary by experience with
international players decreasing slightly earlier relative to college
or high school players.

Table \ref{GameMaxPerformance} provides estimates of the number of
games need to reach peak performance. Generally speaking players
drafted straight out of high school take longer to reach maximum
performance while those with college experience are the quickest. That
said, any conclusions drawn from Table \ref{GameMaxPerformance} or
Figure \ref{careerPrediction} should be made with care as some of the
curves are accompanied with a moderate to substantial amount of variability.

%
\begin{table}[htdp]
\caption{Predicted game at which max performance is attained
(prediction errors are in parenthesis).}\label{GameMaxPerformance}
\vspace*{4pt}
\begin{tabular}{l l ccc}
& &\multicolumn{3}{c}{Age During First Game Played} \\ \cline{3-5}
Draft & Experience & 19 & 21 & 23\\
\hline
\multirow{3}{*}{Top 5} & High School & 477(169.8) & - & - \\
& College & 385(162.2) & 437(184.9) & 386(178.9) \\
& International & 399(155.6) & 448(160.7) & 423(147.1) \\ \cline{1-5}
\multirow{3}{*}{1st Round} & High School & 472(147.4) & - & - \\
& College & 436(185.5) & 417(193.5) & 403(200.2) \\
& International & 446(151.3) & 473(172.4) & 413(167.3) \\ \cline{1-5}
\multirow{3}{*}{2nd Round} & High School & 568(184.7) & - & -\\
& College & 344(152.6) & 386(179.1) & 369(179.7) \\
& International & 409(144.6) & 416(151.1) & 419(169.6) \\
\hline
\end{tabular}
\end{table}
%

%
\begin{figure}[t!]
\includegraphics{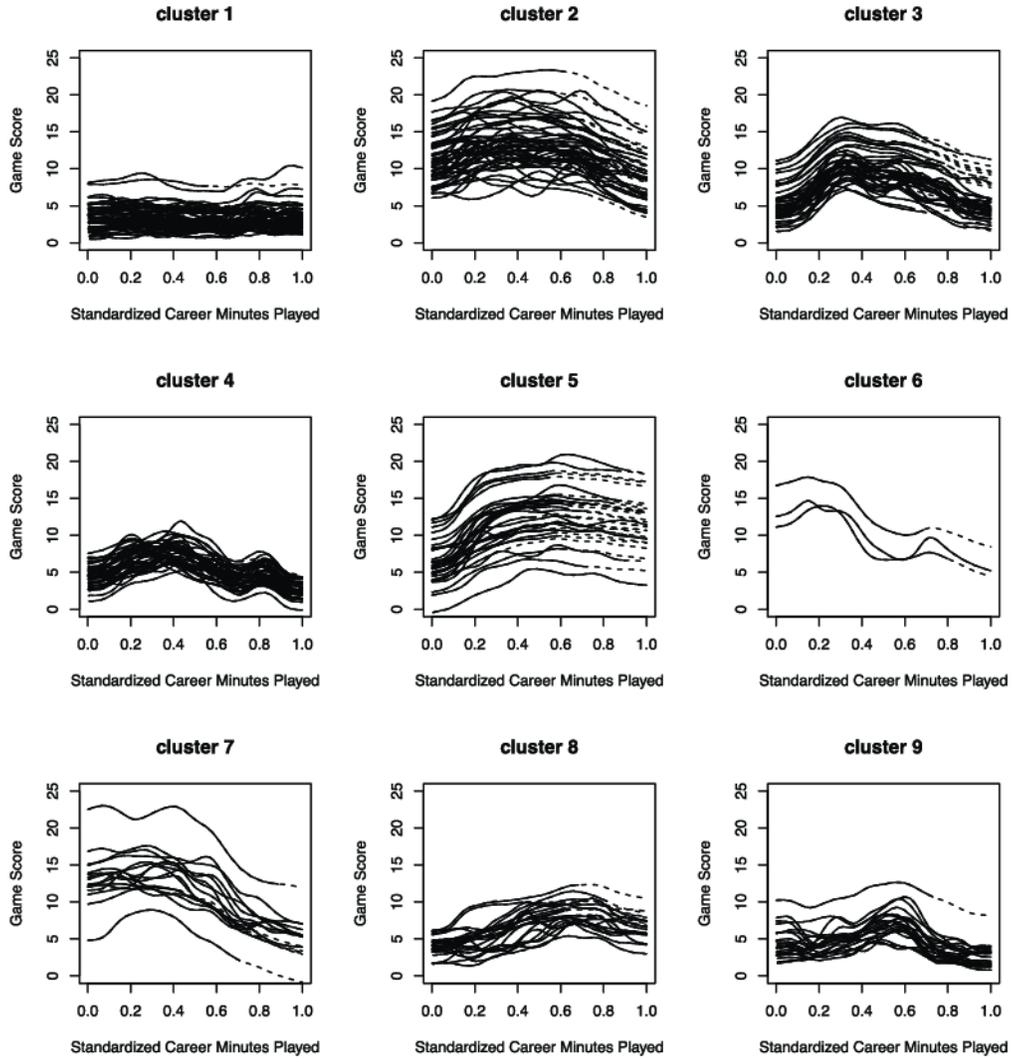}
\caption{Player specific posterior mean production curves divided into
the 9 clusters corresponding to the partition estimated using \cite
{dahl:2006}'s cluster estimate method. Active player prediction for
active players is displayed by a dashed line.}
\label{AllClusterCurves}
\end{figure}

\subsubsection{Cluster Analysis}
A nice property of the model is the ability to postulate what
characteristics guide clustering. To do this it is necessary to obtain
a point estimate using cluster MCMC iterates. Since posterior summaries
of cluster specific parameters are arbitrary, using typical posterior
summaries (mean and median) makes little sense. We employ \cite
{dahl:2006}'s method which is based on least-squares distance from the
matrix of posterior pairwise co-clustering probabilities. Using this
method produces a partitioning of the 408 players into 18 clusters with
cluster membership ranging from 3 to 63 players. Figure \ref
{AllClusterCurves} provides player-specific posterior mean curves for
nine clusters. Except for cluster 6, these clusters represent those
that contain the highest number of players (approximately 80\% of
players). Cluster 6 was selected as it contains curves that are in our
opinion more interesting than clusters not shown. The remaining nine
clusters for the most part are comprised of role players. Although each
cluster contains curves that are slightly unique, they are relatively
flat. The dashed segments at the end of some curves are active player
predictions. To facilitate comparisons we maintain the $x$-axis on the
percentile number of games played scale. We highlight a few of the
clusters by pointing out some of the well known players. Cluster 1 is
comprised of role players (e.g., Tony Allen and Matt Bonner) whose
production is constant. Cluster 2's key member is LeBron James. Players
in this cluster begin careers close to peak level and appear to
maintain production. Cluster 3 contains Carlos Boozer and Ron Artest
who had sharp increase in production but maintained peak performance
for a short time with a gradual decrease in performance. Cluster 4 is
comprised of role players who showed an increase in production right
before retiring (e.g. James Posey). Kobe Bryant is the key player of
Cluster 5 (along with Chauncey Billups, Steve Nash). In this cluster,
players started slow but experienced large sharp increases of
production and maintained it for much of their career. Clusters 6 and 7
are comprised of players who start at peak performance and gradually
decline through out their career with players in Cluster 6 showing a
brief increase towards the end of their career. Grant Hill is a member
of Cluster 6 and Shaquille O'Neal is a member of Cluster 7. Clusters 8
and 9 are primarily comprised of role players with Cluster 8 showing
gradual increase until the later stages of a career. An example is Matt
Barnes. Marcus Camby is member of Cluster 9 and the decrease in
production towards the end of the career is more sharp relative to
Cluster 8. Overall, the clusters contain curves that have distinct
shapes. Information provided in Figure \ref{AllClusterCurves} could
potentially be used to guide NBA personnel decisions regarding contract
length and amount. For example, production for players in 3 begins to
decrease earlier than Cluster 2.

\begin{figure}[htbp]
\includegraphics{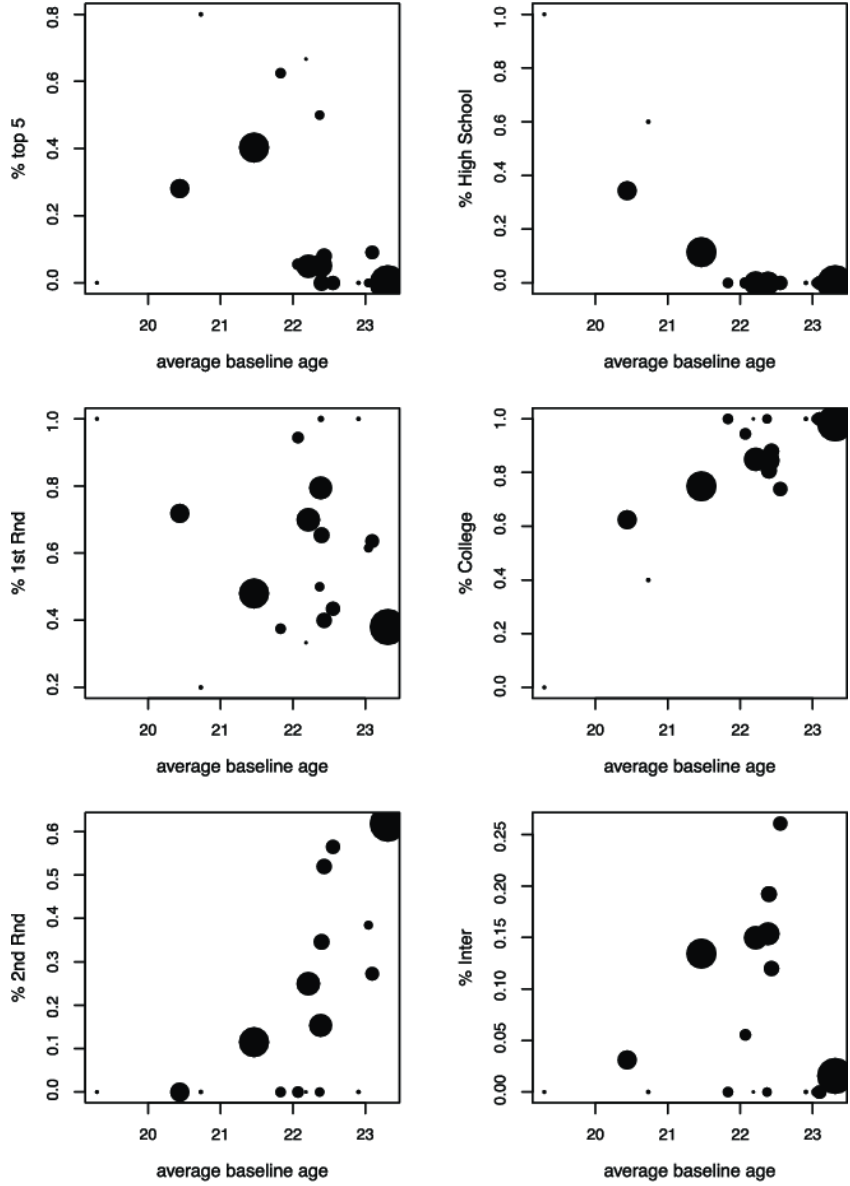}
\caption{Average baseline age and percent of players for each of the 18
clusters for the six categories corresponding to experience and draft
order. The size of the dot is proportional to the number of players
allocated to the cluster.}
\label{ClusInt}
\end{figure}

Finally, Figure \ref{ClusInt} displays the average age during first
game played vs. the percent of players in each of the six categories
for each of the 18 clusters. Apart from demonstrating the presence of
an interaction between the three covariates, these plots confirm what
is already widely known. That is, on average, the age of players
increases as draft order increases and players that play college tend
to begin NBA careers at an older age relative to high school and
international players.

%

\subsection{Assessing Trade-off between Individual Player Fits and Prediction}
As mentioned previously, incorporating the PPMx in the hierarchical
model improves predictions at the cost of a small loss in individual
fits. To show that the cost is minimal relative to gains in prediction,
we randomly selected 50 retired players and removed the final 25\% of
games played (essentially treating them as active players). We then
proceeded to fit four models (details follow) to these partitioned data
and assess model fit through Mean Square Error (MSE). To assess
prediction accuracy, active player prediction was carried out for each
of the 50 randomly selected players and Mean Squared Prediction Error
(MSPE) was computed. The four models considered were the SP model (\ref
{DGM}), an extension of the SP model that improves prediction, the
HPPMx, and the following 5th degree polynomial regression model:
\begin{align*}
y_{it} & = \bm{x}'_i\bm{\beta} + \sum_{j=0}^5 \gamma_{ji}t^j + \epsilon
_{it} \ \ \mbox{with} \ \ \epsilon_{it} \stackrel{iid}{\sim} N(0, \sigma
^2_i) \ \ \mbox{and} \ \ \bm{\beta} \sim N(\bm{0}, s^2\bm{I})\\
\bm{\gamma}_i & \sim N(\bm{\mu}, \bm{T}) \ \ \mbox{where} \ \ \bm{T} =
\mbox{diag}(\tau^2_0, \ldots, \tau^2_5) \\
\bm{\mu} & \sim N(0, s^2\bm{I}).
\end{align*}

The SP model (\ref{DGM}) was extended in the following way
\begin{align*}
\bm{\theta}_i & \sim N(\bm{\mu}, \tau^2\bm{K}^{-1})\\
\bm{\mu} & \sim N(\bm{0}, s^2\bm{I}).
\end{align*}
We refer to this model as hSP (hierarchical semi-parametric).
Predicting (or extrapolating) the last 25\% of games played using the
SP model requires drawing $\theta$'s associated with knots for removed
games from its prior distribution. Therefore, centering the prior on a
vector of global spline coefficients should improve prediction relative
to a prior centered at 0.

The MSE averaged over the 408 players was calculated for each of the
four models resulting in Polynomial (30.72), SP (29.66), hSP (29.71),
HPPMx (31.71). As expected the flexible penalized splines produce the
smallest MSE and HPPMx has the highest MSE illustrating the surrender
of a bit of individual player fit. The MSPE averaged over the 50
randomly selected players turned out to be Polynomial (476.32), SP
(34.00), hSP (31.10), and HPPMx (26.90). As expected HPPMx gained quite
a bit in terms of prediction (extrapolation) accuracy at a fairly
minimal individual fit cost.

\section{Conclusions}

We have proposed a completely novel methodology that incorporates
information regarding the shape of longitudinal curves in predicting
future NBA player production based on age, experience, and draft order
of player. Clearly, curve shape provides information beyond available
covariates and the inclusion of this information in modeling efforts
should improve inferences. In addition, the methodology does well in
balancing individual fits and producing clusters that provide adequate
borrowing of strength among players. The PPMx prior employed does a
nice job of being able to incorporate both covariate and curve shape
information when forming clusters and ultimately borrowing strength
among subjects to improve active player and career predictions. From a
basketball perspective, individual player production clearly depends on
many omitted variables (such as team strength, injury history and
coaching philosophy) and these variables can be easily incorporated in
the model using the PPMx prior when they become available. Finally,
though the methodology was demonstrated using production curves of NBA
basketball players, the idea of incorporating curve shape in inferences
should be applicable in a wide variety of settings (e.g., biomedical,
finance, and environmental studies).

\begin{acknowledgement}
The authors wish to thank the reviewers and editors for their comments,
which improved the paper. The first author was partially funded by
grant FONDECYT 11121131 and the second author was partially funded by
grant FONDECYT 1141057.
\end{acknowledgement}



%


\appendix
\vspace*{-1pt}\section*{Appendix}
\section*{1 John Hollinger's Game Score}

The Hollinger game score is the following linear combination of
statistics appearing in a typical box-score summary of each players
game statistics:

\begin{itemize}
\item PTS = total points scored by player in game\vspace*{-2pt}
\item FGM = number of shots that player made in game\vspace*{-2pt}
\item FGA = number of shots that player attempted in game\vspace*{-2pt}
\item FTM = number of free throws made in game\vspace*{-2pt}
\item FTA = number of free throw attempts in game\vspace*{-2pt}
\item OREB = number of offensive rebounds\vspace*{-2pt}
\item DREB = number of defensive rebounds\vspace*{-2pt}
\item STL = number of steals\vspace*{-2pt}
\item AST = number of assists recorded\vspace*{-2pt}
\item BLK = number of blocked shots recorded\vspace*{-2pt}
\item TO = number of turn overs\vspace*{-2pt}
\item PF = personal fouls.\vspace*{-8pt}
\end{itemize}
\begin{align*}
Game \ Score& =  PTS+ FGM \times0.4 - FGA \times0.7 - (FTA-FTM)\times
0.4 + OREB  \\
&\quad\times0.7 + DREB \times0.3 + STL + AST \times0.7 + BLK \times0.7 - PF \times
0.4\\
&\quad - TO.
\end{align*}

\section*{2 Full Conditionals}
We list the full conditionals used in the Gibbs sampler. In what
follows we use $[\theta|-]$ to denote the distribution of $\theta$
conditioned on all other parameters and data and $n_h$ denotes the
number of subjects belonging to cluster $h$. Also for notational
convience, $\bm{H}_i$ denotes the B-spline basis design matrix for both
$g_i=1$ and $g_1=0$.
\begin{align*}
[\bm{\beta}_{i}|-] & \sim N_m\left( \bm{\mu}_{\beta}, \bm{\Sigma}_{\beta
} \right), \\
& \phantom{{}={}} \bm{\mu}_{\beta} = \left[\sigma^{-2}_i\bm{H}'_{i}\bm
{H}_{i} + \lambda^{-2*}_{s_i}\bm{I}\right]^{-1} \left[ \sigma^{-2}_i\bm
{H}'_{i}(\bm{y}_{i} - \bm{1}_i\beta_{0i}) + \lambda^{-2*}_{s_i}\bm
{\theta}^*_{s_i}\right] \\
& \phantom{{}={}} \bm{\Sigma}_{\beta} = \left[\sigma^{-2}_i\bm
{H}'_{i}\bm{H}_{i} + \lambda^{-2*}_{s_i}\bm{I}\right]^{-1}\\
[\beta_{0i}|-] & \sim N\left(\frac{\sigma^2_{b_0}\sum
_{t=1}^{n_i}[y_{it} - g_i\bm{h}'_{it}\bm{\beta}_{i} - (1-g_i)\tilde{\bm
{h}}'_{it}\bm{\beta}_{i}] + \sigma^2_i\mu_{b_0}}{n_i\sigma^2_{b_0} +
\sigma^2_i}, \frac{\sigma^2_{b_0}\sigma^2_i}{n_i\sigma^2_{b_0} + \sigma
^2_i}\right),\\
[\sigma^2_i|-] & \sim IG\left(0.5n_i + a_0, 0.5\sum_{t=1}^{ _i}[y_{it}
- \beta_{0i} - g_i\bm{h}'_{it}\bm{\beta}_{i} - (1-g_i)\tilde{\bm
{h}}'_{it}\bm{\beta}_{i}]^2 + 1/b_0 \right),\\
[\bm{\theta}^*_h|-] & \sim N_m \Biggl(\left[n_h\lambda^{-2*}_h\bm{I} +
\tau^{-2*}_h\bm{K} \right]^{-1}\left[\sum_{i \in S_h}\lambda^{-2*}_h\bm
{\beta}_i + \tau_{h}^{-2*}\bm{\mu}'\bm{K}\right], \bigl[n_k\lambda
^{-2*}_h\bm{I}\\
&\quad + \tau^{-2*}_h\bm{K} \bigr]^{-1}\Biggr),\\
[\tau^{2*}_h|-] & \sim IG \left(0.5m + a_{\tau}, 0.5(\bm{\theta}^{*}_h
- \bm{\mu})'\bm{K} (\bm{\theta}^{*}_h - \bm{\mu}) + 1/b_{\tau}\right) \\
[\bm{\mu}|-] & \sim N([\bm{K}\sum\tau_h^{-2*} + s^{-2}_{\mu}]^{-1}
[\sum\tau_h^{-2*} \bm{\theta}^*_h \bm{K}], [\bm{K}\sum\tau_h^{-2*} +
s^{-2}_{\mu}]^{-1}) \\
[\mu_{b_0}|-] & \sim N\left(\left[m\sigma^{-2}_{b_0} +
s^{-2}_{b_0}\right]^{-1}[\sigma^{-2}_{b_0}\sum_i \beta_{0i}], \left
[m\sigma^{-2}_{b_0} + s^{-2}_{b_0}\right]^{-1}\right)\\
[\sigma^2_{b_0}|-] & \sim IG(0.5m + a_{b_0}, 0.5 \sum(\beta_{0i} - \mu
_{b_0})^2 + 1/b_{b_0}) \\
[n_i|-] & \sim TN(\eta_i, \delta^2, \tilde{n}_i, \infty)\\
[L_i|-] & \sim TN(\nu_i, \psi^2, \tilde{L}_i, \infty).
\end{align*}

\end{document}